\documentclass[twoside,twocolumn,9pt]{article}
\usepackage{extsizes}
\usepackage[super,sort&compress,comma]{natbib} 
\usepackage[version=3]{mhchem}
\usepackage[left=1.5cm, right=1.5cm, top=1.785cm, bottom=2.0cm]{geometry}
\usepackage{balance}
\usepackage{mathptmx}
\usepackage{sectsty}
\usepackage{graphicx} 
\usepackage{lastpage}
\usepackage[format=plain,justification=justified,singlelinecheck=false,font={stretch=1.125,small,sf},labelfont=bf,labelsep=space]{caption}
\usepackage{float}
\usepackage{fancyhdr}
\usepackage{fnpos}
\usepackage[english]{babel}
\addto{\captionsenglish}{%
  
}
\usepackage{array}
\usepackage{droidsans}
\usepackage{charter}
\usepackage[T1]{fontenc}
\usepackage[usenames,dvipsnames]{xcolor}
\usepackage{setspace}
\usepackage[compact]{titlesec}
\usepackage{hyperref}

\usepackage{threeparttable}
\usepackage{tabularx}
\usepackage{array,multirow} 

\usepackage{epstopdf}

\definecolor{cream}{RGB}{222,217,201}

\begin{document}

\pagestyle{fancy}
\thispagestyle{plain}
\fancypagestyle{plain}{
\renewcommand{\headrulewidth}{0pt}
}

\makeFNbottom
\makeatletter
\renewcommand\LARGE{\@setfontsize\LARGE{15pt}{17}}
\renewcommand\Large{\@setfontsize\Large{12pt}{14}}
\renewcommand\large{\@setfontsize\large{10pt}{12}}
\renewcommand\footnotesize{\@setfontsize\footnotesize{7pt}{10}}
\makeatother

\renewcommand{\thefootnote}{\fnsymbol{footnote}}
\renewcommand\footnoterule{\vspace*{1pt}%
\color{cream}\hrule width 3.5in height 0.4pt \color{black}\vspace*{5pt}} 
\setcounter{secnumdepth}{5}

\makeatletter 
\renewcommand\@biblabel[1]{#1}            
\renewcommand\@makefntext[1]%
{\noindent\makebox[0pt][r]{\@thefnmark\,}#1}
\makeatother 
\renewcommand{\figurename}{\small{Fig.}~}
\sectionfont{\sffamily\Large}
\subsectionfont{\normalsize}
\subsubsectionfont{\bf}
\setstretch{1.125} 
\setlength{\skip\footins}{0.8cm}
\setlength{\footnotesep}{0.25cm}
\setlength{\jot}{10pt}
\titlespacing*{\section}{0pt}{4pt}{4pt}
\titlespacing*{\subsection}{0pt}{15pt}{1pt}

\fancyfoot{}
\fancyfoot[LO,RE]{\vspace{-7.1pt}\includegraphics[height=9pt]{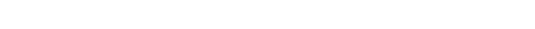}}
\fancyfoot[CO]{\vspace{-7.1pt}\hspace{13.2cm}\includegraphics{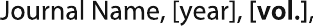}}
\fancyfoot[CE]{\vspace{-7.2pt}\hspace{-14.2cm}\includegraphics{head_foot/RF}}
\fancyfoot[RO]{\footnotesize{\sffamily{1--\pageref{LastPage} ~\textbar  \hspace{2pt}\thepage}}}
\fancyfoot[LE]{\footnotesize{\sffamily{\thepage~\textbar\hspace{3.45cm} 1--\pageref{LastPage}}}}
\fancyhead{}
\renewcommand{\headrulewidth}{0pt} 
\renewcommand{\footrulewidth}{0pt}
\setlength{\arrayrulewidth}{1pt}
\setlength{\columnsep}{6.5mm}
\setlength\bibsep{1pt}

\makeatletter 
\newlength{\figrulesep} 
\setlength{\figrulesep}{0.5\textfloatsep} 

\newcommand{\topfigrule}{\vspace*{-1pt}%
\noindent{\color{cream}\rule[-\figrulesep]{\columnwidth}{1.5pt}} }

\newcommand{\botfigrule}{\vspace*{-2pt}%
\noindent{\color{cream}\rule[\figrulesep]{\columnwidth}{1.5pt}} }

\newcommand{\dblfigrule}{\vspace*{-1pt}%
\noindent{\color{cream}\rule[-\figrulesep]{\textwidth}{1.5pt}} }

\makeatother

\twocolumn[
  \begin{@twocolumnfalse}
{\includegraphics[height=30pt]{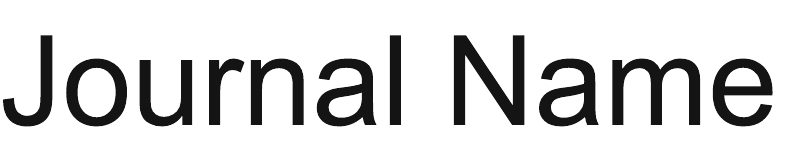}\hfill\raisebox{0pt}[0pt][0pt]{\includegraphics[height=55pt]{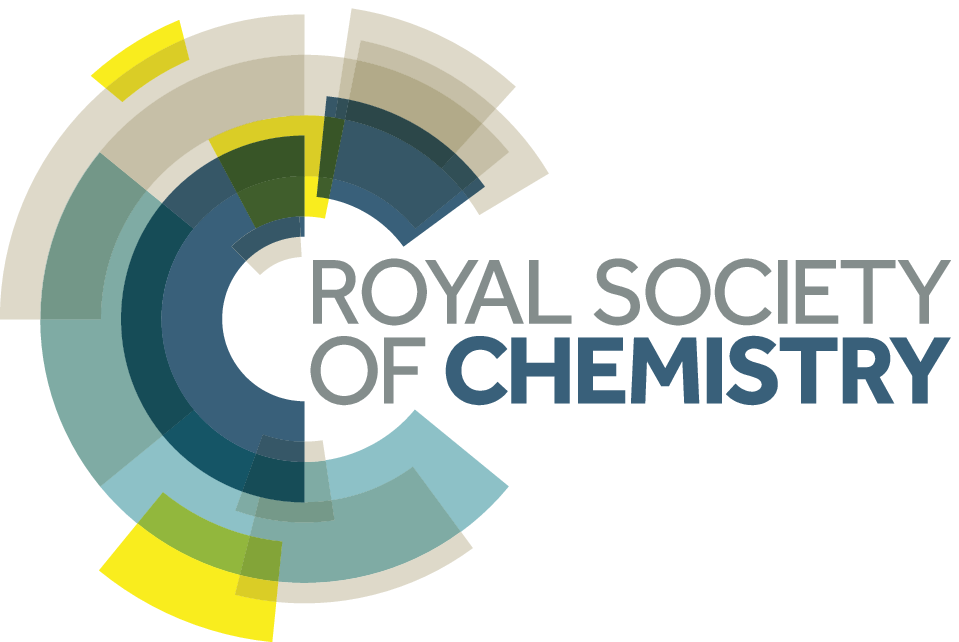}}\\[1ex]
\includegraphics[width=18.5cm]{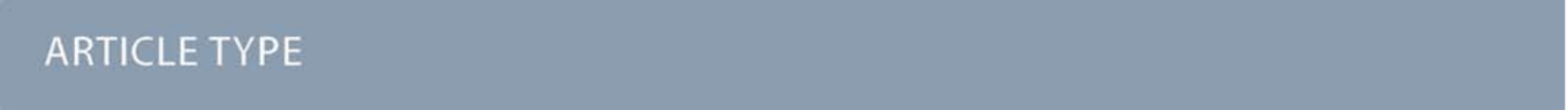}}\par
\vspace{1em}
\sffamily
\begin{tabular}{m{4.5cm} p{13.5cm} }

\includegraphics{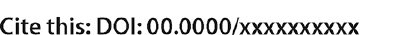} & \noindent\LARGE{\textbf{Suppressing initial capacity fade  in Li-rich  Li$_5$FeO$_4$ with anionic redox by partial Co substitution}} \\
\vspace{0.3cm} & \vspace{0.3cm} \\

 & \noindent\large{Anu Maria Augustine,\textit{$^{a}$}\textit{$^{b}$} Vishnu Sudarsanan,\textit{$^{a}$}\textit{$^{b}$} and P. Ravindran$^{\ast}$\textit{$^{a}$}\textit{$^{b}$}} \\

\includegraphics{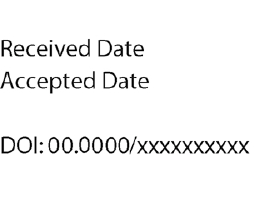} & \noindent\normalsize{The increasing relevance of energy storage technologies demands high-capacity cathode materials for Li-ion batteries. Recently, Li-rich defect anti-fluorite Li$_5$FeO$_4$ has emerged as a high-capacity cathode material exhibiting simultaneous anionic and cationic redox without significant oxygen release. But suffers from irreversible structural change and capacity fade during cycling. Herein we investigate the suppressed capacity fade reported in Co substituted Li$_5$FeO$_4$ and establish the optimal performance through tuning the concentration and oxidation state of Co. We have substituted Co at the Fe site and carried out a detailed analysis of structural, magnetic, electronic, and electrochemical properties using first-principles density functional theory calculations. The extended stability and suppressed capacity fading are found only in specific compositions depending on the Co/Li concentration and the oxidation state of Co. The reasons behind suppressed capacity fading in Co substituted systems have been unveiled on the basis of bonding analyses and proposed as a strategy to suppress the voltage fading reported in Li-rich materials due to transition metal migration. From the evaluation of  thermodynamic stability and electronic structure,  Li$_{5.5}$Fe$_{0.5}$Co$_{0.5}$O$_4$, Li$_{5}$Fe$_{0.25}$Co$_{0.75}$O$_4$, and Li$_{4.5}$Fe$_{0.5}$Co$_{0.5}$O$_4$  are found to exhibit better electrical conductivity than Li$_5$FeO$_4$. All these systems have voltage in the range of 3$-$5 V and exhibit three dimensional Li diffusion pathway with a diffusion barrier height of around 0.3 eV. From Li$_{5}$Fe$_{0.25}$Co$_{0.75}$O$_4$, one can delithiate three Li-ions without structural change and oxygen release, therefore expected to acquire a reversible capacity of around 513 mAhg$^{-1}$. Moreover, as in pristine Li$_5$FeO$_4$, the selected Co substituted systems also exhibit simultaneous anionic and cationic redox without significant O$_2$ release. These results suggest that a tuned Co substitution in Li$_5$FeO$_4$  improves structural stability during delithiation along with promising voltage, electronic and ionic conductivity; eventually suppressing the initial capacity fade and providing high capacity and good kinetics.} \\

\end{tabular}

 \end{@twocolumnfalse} \vspace{0.6cm}

  ]

\renewcommand*\rmdefault{bch}\normalfont\upshape
\rmfamily
\section*{}
\vspace{-1cm}


\footnotetext{\textit{$^{a}$~Department of Physics, Central University of Tamil Nadu, Thiruvarur-610005, India; E-mail: raviphy@cutn.ac.in}}
\footnotetext{\textit{$^{b}$~Simulation Center for Atomic and Nanoscale MATerials, Central University of Tamil Nadu, Thiruvarur, Tamil Nadu, 610005, India. }}




\section{Introduction}
\label{Introduction}
\par

Li-ion battery (LIB) is one of the most successful inventions of this century, , significantly influencing humanity and contributing to the transition towards a green energy future. A vast variety of scientific literatures~\cite{nitta2015li,tarascon1996performance,endo2000recent,edstrom2004cathode,chan2008high,chen2005alpha,lee2010silicon,zhang2004electrochemical,zhang2003low,li2001nanomaterial,liu2009double,liu2012yolk,zhang2011review,doughty2012general,valoen2005transport,zhang2006eis} and a wide range of  applications indicate the keen interest of researchers as well as technologists on improving the performance of LIBs. The growing demand for applications such as portable electronics, electric vehicles, and grid-scale energy storage systems reciprocates the need for improvement in battery technology. Enhancing the energy storage capacity would be regarded as a milestone in the history of the development of batteries and possibly widen the domain of applicability of this technology. Currently, researchers are actively working to achieve the goal of high-capacity batteries~\cite{emani2019li3bn2,zhao20195life0,xiong2020design,ji2019computational,choi2019unexpectedly,shi2018high,zhang2017liquid,nayak2018review,xu2017investigating,yang2017lini0}. The capacity of LIBs greatly depends on the redox potential of the cathode material as well as the number of Li ions that can be exchanged. The traditional lithium transition metal oxide cathodes like LiCoO$_2$ and LiFePO$_4$ can achieve a capacity of fewer than 200 mAhg$^{-1}$ ~\cite{xiong2020design} due to the limited number of Li available for the redox process, which is insufficient to meet the current specific capacity requirements.
\par
Recently, Li-rich materials gained significant popularity due to their ability to deliver a capacity higher than 200 mAhg$^{-1}$~\cite{li2017anionic} by the transfer of multiple Li-ions. Usually, the Li-ion intercalation/de-intercalation process is governed by the redox capacity of the transition metal (TM) in the cathode materials~\cite{luo2016charge}. The capacity of the cathode is limited by the possible oxiation states of the TM as the Li intercalation process involves change in TM's oxidation states. But some of the cathode materials, e.g., Li$_{1.2}$Ni$_{0.13}$Mn$_{0.54}$Co$_{0.13}$O$_2$ deliver a storage capacity of around 300 mAhg$^{-1}$~\cite{assat2017decoupling}, which cannot be solely explained on the basis of TM redox during Li exchange. This remained unresolved until the discovery of anionic redox in Li-rich layered TM oxides~\cite{luo2016charge,sathiya2013reversible,mccalla2015visualization,mccalla2015understanding,sathiya2015electron,grimaud2016anionic,seo2016structural,pearce2017evidence,saubanere2016intriguing,xie2017requirements}. The prediction and verification of anionic redox not only revealed the reason behind extra capacities but also rejuvenated the search for high-capacity cathodes by opening a new direction~\cite{zheng2018mechanism}. As a result, the cathode materials with both anionic and cationic redox that could mitigate the capacity limits became a research hot-spot~\cite{li2017anionic,zheng2018mechanism,lee2015new,zhan2017enabling,grayfer2017anionic,perez2017approaching,yabuuchi2015high,li2019identifying,yu2019revealing,li2019probing,li2019tuning}. But, most of such materials are severely lacking structural stability and therefore end up with poor cycling performance.\\ 

\par
Recent works indicate that anionic redox is not limited to the layered TM oxides alone. The anionic redox mechanism responsible for increasing energy storage capacity has been observed in  cathode materials such as Li$_4$Mn$_2$O$_5$~\cite{freire2016new}, Li$_5$FeO$_4$~\cite{zhan2017enabling}, and some metal-organic compounds~\cite{fang2017metal,zheng2018mechanism}. The materials with an anti-fluorite crystal structure are expected to ease the Li-migration~\cite{hirano2005electrochemical} since such materials are formed in such a way that the Li-ions occupy the oxygen sites, for example, doped zirconia, which is a well-known ion-conducting oxide~\cite{minh1993ceramic}. The compounds such as  Li$_6$CoO$_4$, Li$_6$MnO$_4$, and Li$_5$FeO$_4$ are some of the previously studied Li-rich cathode materials with defect anti-fluorite structure ~\cite{narukawa1999anti,imanishi2005antifluorite,noh2012role,lim2015anti,kirklin2014high,maroni2013characterization,li2019identifying,hoang2018electronic,esaka1987lithium,liang2013synthesis,ding2011solid}. Among these compounds, Li$_5$FeO$_4$ is found to be more attractive as a high-capacity cathode material exhibiting simultaneous anionic and cationic redox activity. Furthermore, the presence of abundantly available 3{\it d} Fe in place of 4{\it d} and 5{\it d} TMs like Ru and Ir in common Li-rich materials ~\cite{pearce2020anionic,nagao2019fast} has a beneficial impact on the molecular mass and the cost of the material. The Li$_5$FeO$_4$ compound with defect anti-fluorite structure was first proposed by Demoisson {\it et al.} in  1971~\cite{demoisson1971antifluorite,stewner1971struktursystematik} and first reported as a promising cathode material by Takeda in 2002~\cite{imanishi2005antifluorite}.  Recently, Zhan {\it et al.}~\cite{zhan2017enabling} have extensively investigated its electrochemical properties, especially the structural changes, voltage profile, and oxygen redox, using both computational and experimental approaches. It was revealed that the Li$_5$FeO$_4$ undergoes reversible anionic and cationic redox with the lithiation/delithiation of two Li-ions. The high degree of delithiation results in the coordination change of Li and Fe from tetrahedral to octahedral (LiO$_4$$ \rightarrow$ LiO$_6$, FeO$_4$$ \rightarrow$ FeO$_6$),--), consequently altering the structure from defect antifluorite to distorted rock salt. This change is permanent, and the initial capacity will be lost in subsequent cycling~\cite{zhan2017enabling}. The experimental investigation by Imanishi {\it et al.}~\cite{imanishi2005antifluorite} found that this tendency is less evident in Co substituted Li$_5$FeO$_4$ and therefore can reversibly delithiate more Li-ions in the Co substituted system compared to the pristine Li$_5$FeO$_4$ material. There are experimental studies on the Co-substituted Li$_5$FeO$_4$ evidencing the favorable electrochemical characteristics~\cite{imanishi2005antifluorite,ding2011solid}. However, there is an apparent lack of detailed study on the dependence of Co valency change and Li/Co concentration on the electrochemistry, including voltage change and Li-diffusion mechanism in Li$_5$FeO$_4$. 

\par 
So, in the present study, we have investigated the structural, electronic, and electrochemical properties of Co substituted Li$_5$FeO$_4$ using density functional theory (DFT) calculations. Since the Co atom can have multiple oxidation states such as +2, +3, and +4 depending upon the chemical environment, we have made attempts to control the charge on the Co atoms by changing the number of Li-ions in its vicinity and verified the oxidation states of Co through various charge analyses schemes. We have also investigated the thermal stability, electronic structure, and bandgap of all these materials. Moreover, along with Li$_5$FeO$_4$, the following Co substituted systems such as Li$_{5.5}$Fe$_{0.5}$Co$_{0.5}$O$_4$, Li$_{5}$Fe$_{0.25}$Co$_{0.75}$O$_4$, and Li$_{4.5}$Fe$_{0.5}$Co$_{0.5}$O$_4$ are shortlisted for further analyses considering their low bandgap values which indicates their improved electrical conductivity. The following sections contain detailed analyses of electronic structure, chemical bonding, Li diffusion kinetics, and voltage profile of these compounds. Finally, we have discussed the electrochemistry behind the delithiation of these materials, such as cationic and anionic redox and oxygen release, to have a conclusive picture of Co substitution in Li$_5$FeO$_4$  to use them as cathode material for high capacity Li-ion batteries.

\section{Computational details}
\label{Computational details}
\par
The present study is carried out using DFT calculations as implemented in the Vienna {\it ab-initio} simulation package (VASP)~\cite{kresse1996efficient,kresse1996efficiency}. The projector augmented wave (PAW) potentials~\cite{blochl1994projector} are used to replace the core electrons. The generalized gradient approximation (GGA) formulated by Perdew, Burke, and Ernzerhof~\cite{perdew1996generalized}  with its GGA + {\it U}~\cite{dudarev1998electron} extension is used to treat the exchange-correlation interactions of electrons. The {\it U} value used to account for the correlation effect of {\it d} electrons for Co and Fe is 3.2 and 4.0 eV, respectively~\cite{zhan2017enabling}. 

\par
All the first-principles calculations have been carried out using a plane-wave basis set with an energy cut-off of 520\,eV and a total energy convergence threshold of 10\textsuperscript{-6} eV/unit cell is used for the electronic energy minimization step. The Brillouin zone has been sampled using Monkhorst Pack (MP) technique using a {\bf k}-grid size of 4 x 4 x 4, in accordance with the larger unit cell size. For the density of states (DOS) calculations, a relatively higher {\bf k}-grid size of 6 x 6 x 6 is used. The activation energy calculations for the Li-ion diffusion have been carried out using the climbing-image nudged elastic band (CI-NEB) method as implemented in VASP transition state tools (VTST).~\cite{henkelman2000climbing}  We have used seven intermediate images for each CI-NEB calculation with force minimization of 0.05 eV \AA\textsuperscript{-1} for the atoms perpendicular to the direction of the reaction path as threshold criteria for convergence. The average voltage, as well as voltage profile, have been calculated using GGA + {\it U} method and non-empirical strongly constrained and appropriately normed (SCAN) meta-GGA functional with its SCAN + {\it U} extension~\cite{sun2015strongly}. The delithiated configurations were enumerated for convex-hull construction using the clusters approach to statistical mechanics (CASM) package~\cite{thomas2013finite}.  The formation energies of all the enumerated configurations were calculated from the DFT calculations with complete geometry optimization.

\section{Results and discussion}
\label{Results and discussion}

\subsection{Structural details}
\label{Structural details}

\begin{figure*}[h]
\centering
\includegraphics[scale=0.6]{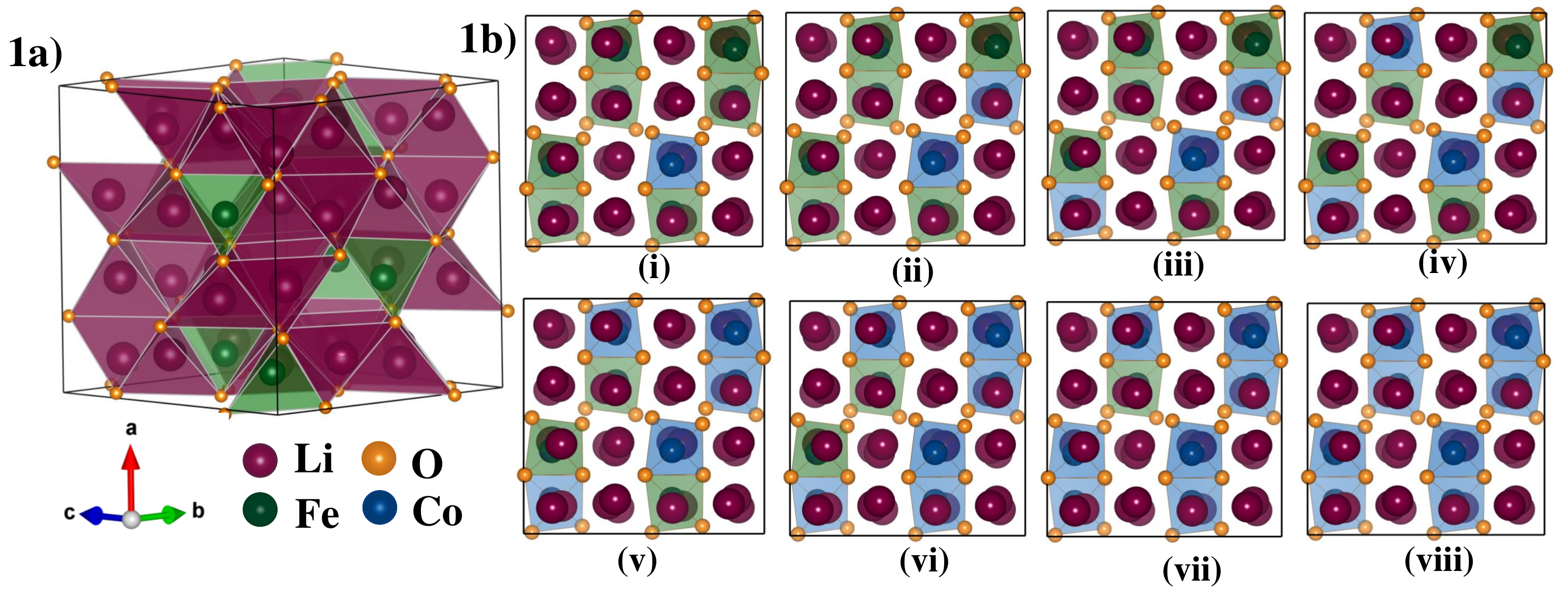}
\caption{Crystal structures of (a) Li$_{5}$FeO$_{4}$ where the FeO$_4$ and LiO$_4$ tetrahedra are given in polyhedral representation and (b) its Co substituted derivatives (Li$_5$Fe$_{1-x}$Co$_x$O$_4$) in the [100] view. The concentration of Co atom increases from (i) to (viii) ( $x$ = (i) 0.125, (ii) 0.250, (iii) 0.375, (iv) 0.500, (v) 0.625, (vi) 0.750, (vii) 0.875, (viii) 1.00 )}
\label{fig:1}
\end{figure*}

\par 
The crystal structure of a material is one of the fundamental selection criteria for battery cathodes. A promising battery cathode based on an intercalation mechanism should be able to easily lithiate/delithiate without structural collapse. Moreover, the electrochemical potential of a system is directly related to the energy needed to add/remove Li from the lattice site and the energy required to reduce/oxidize the cations. These energies are structure-dependent. So in this section, we are analyzing the structural details of  Li$_5$FeO$_4$  and its Co substituted derivatives. At ambient conditions, Li$_5$FeO$_4$ exists in an orthorhombic crystal structure with space group {\it Pbca} and it is usually referred as a defect anti-fluorite structure. This structure can be defined as the substitution of a single  Fe$^{3+}$ ion by removing three Li from the 2 x 2 x 2 supercell of rock-salt phase Li$_2$O (space group  {\it Fm$\overline{3}$m} ). This substitution will introduce 16 Li vacancies in the crystal lattice. These vacant interstitial sites provide a low-energy pathway for the Li-ion diffusion~\cite{johnson2010li2o}. In Li$_5$FeO$_4$, each Li and Fe ion is tetrahedrally coordinated by O as shown in Fig.~\ref{fig:1}(a). 

\par
In the Co substituted systems, we have replaced Fe ion in the parent structure with Co ion sequentially, preserving the homogeneity of distribution of substituent as shown in Fig.~\ref{fig:1}(b). The number of Li ions near the substituent has been adjusted to monitor the charge on Co ions, i.e., a Li-ion close to the substituted Co has been removed to obtain a higher charge state on Co sites. In contrast, if an extra Li has been introduced at the previously vacant Li site in the proximity of Co, the charge state of Co ions will be lowered. We have also made a pure substitution of Fe with Co, of which the Li concentration is left undisturbed. In this case, we expect that the Co ions will have the same charge state as Fe (i.e., +3) in Co substituted Li$_5$FeO$_4$. Complete structural optimization of the crystal structure of Co substituted systems are performed and the corresponding optimized equilibrium lattice parameters are listed in Table\,\ref{tab:1}. 

\begin{table*}[]
\centering
\caption{The optimized equilibrium structural parameters and magnetic moments from ferromagnetic calculations  for Li$_{y}$Fe$_{(1-x)}$Co$_{x}$O$_4$ systems where {\it y = 5+x}, {\it y = 5} and {\it y = 5-x}; $x = 0.00$, 0.125, 0.250, 0.375, 0.500, 0.625, 0.750, 0.875, 1.00)}
\begin{threeparttable}
\label{tab:1}
\begin{tabularx}{465pt}{ m{4cm} X X X X X X X }
\hline 
\multirow{2}{*}{Compound} & \multicolumn{4}{c}{ structural parameters} &  \multicolumn{3}{c}{ magnetic moment} \\\cline{2-8}
     & a ({\AA}) &  	b ({\AA}) &	c ({\AA}) & 	V ({\AA}\textsuperscript{3}) & Fe ($\mu_B$/atom) & Co ($\mu_B$/atom)&  total ($\mu_B$/f.u.) \\ 
\hline
Li$_5$FeO$_4$ & 9.195 & 9.253 & 9.244 & 786.43  &  4.116  & & 4.796\\[0ex]
 & 9.153\tnote{{\it a}} & 9.218\tnote{{\it a}} & 9.213\tnote{{\it a}} & 777.32\tnote{{\it a}} & & &  \\
Li$_{5.125}$Fe$_{0.875}$Co$_{0.125}$O$_4$ & 9.243  & 9.244 & 9.258 & 791.01 & 4.117 & 2.586 & 4.554\\
Li$_{5.25}$Fe$_{0.75}$Co$_{0.25}$O$_4$ & 9.292 & 9.239 & 9.268 & 795.70 & 4.117 & 2.586 & 4.313 \\
Li$_{5.375}$Fe$_{0.625}$Co$_{0.375}$O$_4$ & 9.332  & 9.226 & 9.282 & 799.19 & 4.115 & 2.574 & 4.072 \\
Li$_{5.5}$Fe$_{0.5}$Co$_{0.5}$O$_4$ & 9.342 & 9.265 & 9.290  & 804.12 & 4.121 & 2.588 & 3.830\\
Li$_{5.625}$Fe$_{0.375}$Co$_{0.625}$O$_4$ & 9.336 & 9.267 & 9.316 & 805.84 & 4.121 & 2.588 & 3.588 \\
Li$_{5.75}$Fe$_{0.25}$Co$_{0.75}$O$_4$ & 9.331 & 9.285 & 9.339 & 809.13 & 4.124& 2.588 & 3.348 \\
Li$_{5.875}$Fe$_{0.125}$Co$_{0.875}$O$_4$ & 9.347 & 9.263 & 9.348 & 809.41 & 4.134 & 2.591 & 3.107 \\ 
Li$_6$CoO$_4$ & 9.351 & 9.305 & 9.361 & 814.41 & & 2.594 & 2.866  \\
Li$_{5}$Fe$_{0.875}$Co$_{0.125}$O$_4$ & 9.194 & 9.248 & 9.232 & 784.97 & 4.116 & 2.953 & 4.668 \\
Li$_{5}$Fe$_{0.75}$Co$_{0.25}$O$_4$ & 9.196 & 9.239 & 9.222 & 783.51 & 4.116 & 2.954 & 4.540 \\
Li$_{5}$Fe$_{0.625}$Co$_{0.375}$O$_4$ & 9.196 & 9.226 & 9.209 & 781.26 & 4.117 & 2.952 & 4.412 \\
Li$_{5}$Fe$_{0.5}$Co$_{0.5}$O$_4$ & 9.191 & 9.219 & 9.206 &  780.11 & 4.116 & 2.951 & 4.285 \\
Li$_{5}$Fe$_{0.375}$Co$_{0.625}$O$_4$ & 9.192 & 9.204 & 9.191 & 777.57 & 4.116 & 2.953 & 4.156\\
Li$_{5}$Fe$_{0.25}$Co$_{0.75}$O$_4$ & 9.197 & 9.190 & 9.179 & 775.85 & 4.116 & 2.953 & 4.028\\
Li$_{5}$Fe$_{0.125}$Co$_{0.875}$O$_4$ & 9.195 & 9.178 & 9.167 & 773.60 & 4.116 & 2.952 & 3.901 \\ 
Li$_5$CoO$_4$ & 9.194 & 9.175 & 9.153 & 772.12 & & 2.953 & 3.773 \\
Li$_{4.875}$Fe$_{0.875}$Co$_{0.125}$O$_4$ & 9.174 & 9.261 & 9.243 & 785.23 & 4.115 & 3.143 & 4.777\\
Li$_{4.75}$Fe$_{0.75}$Co$_{0.25}$O$_4$ & 9.163 & 9.275 & 9.240 & 785.30 & 4.113 & 3.143 & 4.758\\
Li$_{4.625}$Fe$_{0.625}$Co$_{0.375}$O$_4$ & 9.154 & 9.280 & 9.236 & 784.54 & 4.113 & 3.145 & 4.740 \\
Li$_{4.5}$Fe$_{0.5}$Co$_{0.5}$O$_4$ & 9.125 & 9.288 & 9.226 &  781.88 & 4.113 & 3.148 & 4.722 \\
Li$_{4.375}$Fe$_{0.375}$Co$_{0.625}$O$_4$ & 9.105 & 9.291 & 9.238 & 781.42 & 4.114 & 3.146 & 4.703 \\
Li$_{4.25}$Fe$_{0.25}$Co$_{0.75}$O$_4$ & 9.095 & 9.270 & 9.258 & 780.45 & 4.114 & 3.146 & 4.685\\
Li$_{4.125}$Fe$_{0.125}$Co$_{0.875}$O$_4$ & 9.124 & 9.227 & 9.240 & 777.85 & 4.114 & 3.147 & 4.668 \\ 
Li$_4$CoO$_4$ & 9.090 & 9.310 & 9.179 & 776.71 &  & 3.150 &  4.647 \\
\hline
\end{tabularx}
\begin{tablenotes}
\item[{\it a}]{Ref.~\cite{johnson2010li2o}} 
\end{tablenotes}
\label{tab:1}

\end{threeparttable}
\end{table*}

\par

\begin{figure}[h]
\centering
\includegraphics[scale=0.8]{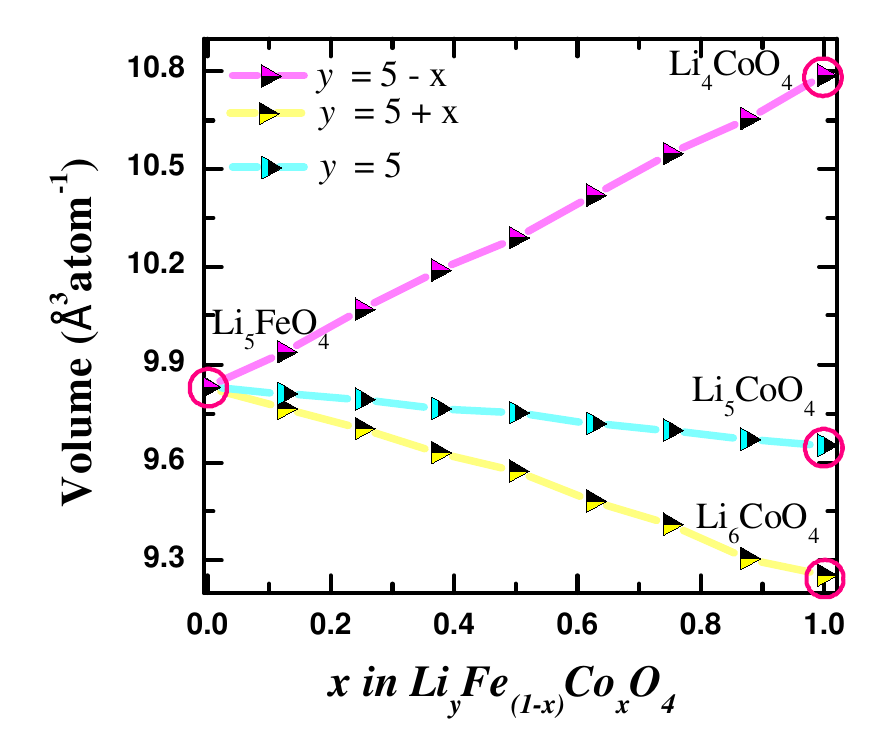}
\caption{The change in volume per atom as a function of Co concentration ($x$) for various Li concentrations ($y$) in Li$_{y}$Fe$_{(1-x)}$Co$_x$O$_4$. The end member compositions are given in the corresponding curves }
\label{fig:2}
\end{figure}

 In order to have a high volumetric capacity, the cathode material should have a maximum number of intercalating Li-ions within a lower volume. Moreover, the volume change during the lithiation/delithiation process will significantly impact the cycling stability. A large volume change upon charging/discharging may reduce the electrical contact of the electrode with current collectors and occasionally damage the electrodes by micro-crack formation. So we have extensively investigated the volume change in Li$_{y}$Fe$_{(1-x)}$Co$_x$O$_4$ with Co substitution as well as Li concentration change. The volume change upon the substitution of Co in +2, +3, and +4 oxidation states is presented in Table~\ref{tab:1} and Fig.~\ref{fig:2}. Table~\ref{tab:1} displays the equilibrium volume per unit cell  by varying Li as well as Co concentration of  Li$_{y}$Fe$_{(1-x)}$Co$_{x}$O$_4$, while Fig.~\ref{fig:2} represents the volume per atom for the corresponding systems.  In Li$_{5}$Fe$_{(1-x)}$Co$_x$O$_4$, the increase in the concentration of Co substitution  results in a decrease in volume per unit cell and a corresponding decrease in volume per atom, as evident from Table~\ref{tab:1}  and Fig.~\ref{fig:2}. On the other hand, for the series of systems like Li$_{(5+x)}$Fe$_{(1-x)}$Co$_x$O$_4$ and Li$_{(5-x)}$Fe$_{(1-x)}$Co$_x$O$_4$, the volume is affected by not only the Co substitution but also the accompanied addition or removal of Li. It can be expected that the removal or addition of Li is more influential in the volume change than the Co substitution as the Li addition/removal changes the number of atoms in the unit cell. In contrast, Co substitution will not change the total number of atoms in these systems. It can be found from Table~\ref{tab:1} that  the volume per unit cell increases when one increase the Co$^{+2}$ concentration which is accompanied by the addition of Li (i.e. Li$_{(5+x)}$Fe$_{(1-x)}$Co$_x$O$_4$). On the other hand, the Co$^{+3}$ substitution with the removal of Li (i.e. Li $_{(5-x)}$Fe$_{(1-x)}$Co$_x$O$_4$) results in a decrease in the volume per unit cell. However, the volume change is not as pronounced as one would anticipate. The Li addition or removal is not significantly impacting the volume even though the effect is greater than that from Co substitution. This property can be attributed to the characteristic of the crystal structure of Li$_5$FeO$_4$. As discussed before, Li$_5$FeO$_4$ contains 16 Li vacant sites in its unit cell. Such lower volume change by the lithiation/delithiation process is associated with many Li vacant sites available to accommodate Li into this system. Further confirmations can be found from Fig.~\ref{fig:2} that, though the volume per atom increase in Li$_{(5-x)}$Fe$_{(1-x)}$Co$_x$O$_4$ with Co substitution,  it decrease in Li$_{(5+x)}$Fe$_{(1-x)}$Co$_x$O$_4$. Moreover, observing a small volume change upon Li addition/removal indicates that the volume change associated with the lithiation/delithiation process during the cycling will also be small. This is an essential property for a good cathode material, as discussed above. It may also be noted that the removal of Li will create additional Li vacant sites in the unit cell, which is expected to make Li diffusion easier than the Li$_5$FeO$_4$.

We have performed additional total energy calculations to verify that the extra Li which accompanies the Co substitution in Li$_{5+x}$Fe$_{1-x}$Co$_x$O$_4$ systems, actually prefers to reside near the Co site rather than the Fe site. Complete structural optimization of ferromagnetic  Li$_{5.125}$Fe$_{0.875}$Co$_{0.125}$O$_4$ in these two scenario is performed. One with Li inserted to the vacant site near Fe and another with Li inserted to the vacant site near to Co.  From these calculations, we have found that a Li to occupy a vacant site close to Fe requires 17\,meV more energy than occupying a vacant site closer to Co. This indicates that the Li ions prefer to occupy the vacant site closer to Co in Li$_{y}$Fe$_{(1-x)}$Co$_{x}$O$_4$. So we have done all our calculations accordingly.

\subsection{Magnetic structure of  Li$_{y}$Fe$_{(1-x)}$Co$_{x}$O$_4$ }
\label{Magnetic structure}
In the present study, we have investigated the magnetic ordering in Li$_5$FeO$_4$ using first-principles calculations. To find the ground-state magnetic ordering of Li$_5$FeO$_4$ we have considered the nonmagnetic,  ferromagnetic, {\it A}, {\it C}, and {\it G}-type antiferromagnetic orderings. From these calculations, it has been found that the Li$_5$FeO$_4$  exhibits an {\it A}-type antiferromagnetic ordering in its ground state. The ferromagnetic ordering is only 1.9\,mev/f.u higher in energy than the ground state {\it A}-type antiferromagnetic ordering. The nonmagnetic ordering results 3.2\,eV/f.u higher energy than the {\it A}-type antiferromagnetic ground state. Also, we have carried out ferromagnetic and anti-ferromagnetic ({\it A, C}, and {\it G}-type) calculations for Li$_{5}$Fe$_{0.25}$Co$_{0.75}$O$_4$. From this study, we found that the ferromagnetic ordering is the lowest energy configuration over the antiferromagnetic orderings.  But, the {\it C}-type anti-ferromagnetic ordering has just 6\,meV higher energy per formula unit than that of the ferromagnetic ground state.  This indicates that for certain Li and Co concentrations and thermodynamic conditions, the Li$_{y}$Fe$_{(1-x)}$Co$_{x}$O$_4$ system may have antiferromagnetic ordering in the ground state as in the pristine Li$_5$FeO$_4$. But, for computational simplicity, ferromagnetic structural ordering is considered for all the systems in the present study to account for the spin polarization effect in the calculations.

The calculated magnetic moment of Fe and Co atoms in  Li$_{y}$Fe$_{(1-x)}$Co$_{x}$O$_4$  systems from our ferromagnetic calculations are given in Table~\ref{tab:1}. In Li$_{y}$Fe$_{(1-x)}$Co$_{x}$O$_4$,  Fe is expected to have the +3 oxidation state. The Fe$^{3+}$ ion in the high spin (HS) state will have the magnetic moment of 5\,$\mu_B$/atom if we consider the pure ionic picture.  By analyzing the magnetic moments, we found that Fe has a magnetic moment of around 4.1\,$\mu_B$/atom and this indicates that Fe$^{3+}$ is in the HS state and the magnetic moment is quenched due to the presence of covalency between Fe$-$O. The magnetic moment at the Fe site is not changed significantly irrespective of the chemical environment and this suggests that the Fe is always in +3 oxidation with the HS state. The Co$^{2+}$ ion in HS is expected to have the magnetic moment of 3\,$\mu_B$/atom. However, in Li$_{(5+x)}$Fe$_{(1-x)}$Co$_{x}$O$_4$, the Co exhibits a magnetic moment around 2.6\,$\mu_B$/atom suggests that Co is in +2 oxidation state with HS state and the covalency reduced the magnetic moment from 3 to 2.58\,$\mu_B$/atom. The HS Co ion in the +3 and +4 oxidation states are expected to have a magnetic moment of 4 and 5 \,$\mu_B$/atom, respectively. So, in  Li$_{5}$Fe$_{(1-x)}$Co$_{x}$O$_4$ and Li$_{(5-x)}$Fe$_{(1-x)}$Co$_{x}$O$_4$ we are getting  the magnetic moment at the Co site of  2.95 and 3.14\,$\mu_B$/atom respectively. However, when we go from +2 to +3 to +4 oxidation state of Co ion, the covalency between Co$-$O bond systematically increases and hence the Co$^{4+}$ ion with HS state magnetic moment reduced from 5\,$\mu_B$/atom (pure ionic picture) to 3.14\,$\mu_B$/atom. The present observation of increasing the covalency with increase of oxidation state of Co is consistent with the conclusion arrived from our ICOHP analysis.  This increase in covalency is the key factor for drawing out more Li from Co substituted systems without destabilizing the structure compared with pure Li$_5$FeO$_4$.

\subsection{Enthalpy of formation}
\label{Enthalpy of formation}

\par

The enthalpy of formation of a material is a primary property that can give information about its thermal stability. If the material has a negative value of formation enthalpy, then one can conclude that the material will be thermally stable and possible to synthesize. A positive value of enthalpy of formation indicates that such compounds cannot be stabilized at ambient conditions. The enthalpy of formation($\Delta H\textsubscript{f}$) can be calculated using the following formula:

\begin{equation} \label{eu_eqn1}
\Delta H\textsubscript{f}\textsuperscript{Li$_y$Fe$_{1-x}$Co$_x$O$_4$} = E\textsubscript{t}\textsuperscript{Li$_y$Fe$_{1-x}$Co$_x$O$_4$}-(yE\textsubscript{t}\textsuperscript{Li}+(1-x)E\textsubscript{t}\textsuperscript{Fe}+xE\textsubscript{t}\textsuperscript{Co}+4E\textsubscript{t}\textsuperscript{O} )
\end{equation}

where E\textsubscript{t}\textsuperscript{Li$_y$Fe$_{1-x}$Co$_x$O$_4$}, E\textsubscript{t}\textsuperscript{Li}, E\textsubscript{t}\textsuperscript{Fe}, E\textsubscript{t}\textsuperscript{Co}, and E\textsubscript{t}\textsuperscript{O} are the total energy per formula unit for respective systems obtained for the equilibrium geometry through structural optimization. The enthalpy of formation of all the systems considered in the present study are calculated using Equation:\ref{eu_eqn1} and are listed in Table~\ref{tab:2}.

\begin{table*}[]
\centering
\caption{Calculated enthalpy of formation for  Li$_{y}$Fe$_{(1-x)}$Co$_{x}$O$_4$ systems (where {\it y = 5+x}, (b) {\it y = 5} and (c) {\it y = 5-x} ; $x = 0.00$, 0.125, 0.250, 0.375, 0.500, 0.625, 0.750, 0.875, 1.00)}
\begin{threeparttable}
\label{tab:2}
\begin{tabularx}{410pt}{ l c c c c c c c c c }
\hline
\multirow{2}{*}{Compound} & \multicolumn{9}{c}{ $\Delta H\textsubscript{f}\textsuperscript{Li$_y$Fe$_{1-x}$Co$_x$O$_4$}$ (eV/atom)}   \\\cline{2-10}

  & x =  0.00 & 0.125 & 0.250 & 0.375 & 0.500 & 0.625 &  0.750 & 0.875 & 1.00 \\ 
\hline
Li$_{(5+x)}$Fe$_{(1-x)}$Co$_{x}$O$_4$  &  & -1.91 & -1.88 & -1.86  & -1.84 & -1.82 & -1.81 & -1.79 & -1.78 \\
Li$_{5}$Fe$_{(1-x)}$Co$_{x}$O$_4$  & -1.94 & -1.92 & -1.90 & -1.88 & -1.87 & -1.85 & -1.83 & -1.81 & -1.80 \\
Li$_{(5-x)}$Fe$_{(1-x)}$Co$_{x}$O$_4$ &  & -1.90 & -1.86 & -1.83 & -1.80 & -1.76 & -1.72 & -1.69 & -1.66\\
\hline
\end{tabularx}
\label{tab:2}
\end{threeparttable}
\end{table*}

\par
The calculated values of $\Delta H\textsubscript{f}$ in Table~\ref{tab:2} indicate that all these compositions are thermally stable with the parent system being the most stable among all the compositions considered here, i.e., both the Co substitution and Li extraction/addition in Li$_5$FeO$_4$ are reducing the phase stability. A comparative analysis of $\Delta H\textsubscript{f}$ shows that the gradual Li additions (Li$_{5+x}$Fe$_{1-x}$Co$_x$O$_4$) reduces the stability slower than the Li extractions (Li$_{5-x}$Fe$_{1-x}$Co$_x$O$_4$) in Li$_5$FeO$_4$. This variation may possibly be due to the preference of Co to have +2 or +3 over +4 oxidation states. Even though such changes in concentration of Li as well as Co in Li$_5$FeO$_4$  decrease the structural phase stability, none of the considered compositions are found to be unstable (positive $\Delta H\textsubscript{f}$).

\subsection{Electronic structure and chemical bonding analyses}
\label{ Electronic structure and chemical bonding analyses}

\subsubsection{The total and projected density of states analyses}
\label{The total and projected density of states analyses}

\par
\begin{figure*}[h]
\centering
\includegraphics[scale=1]{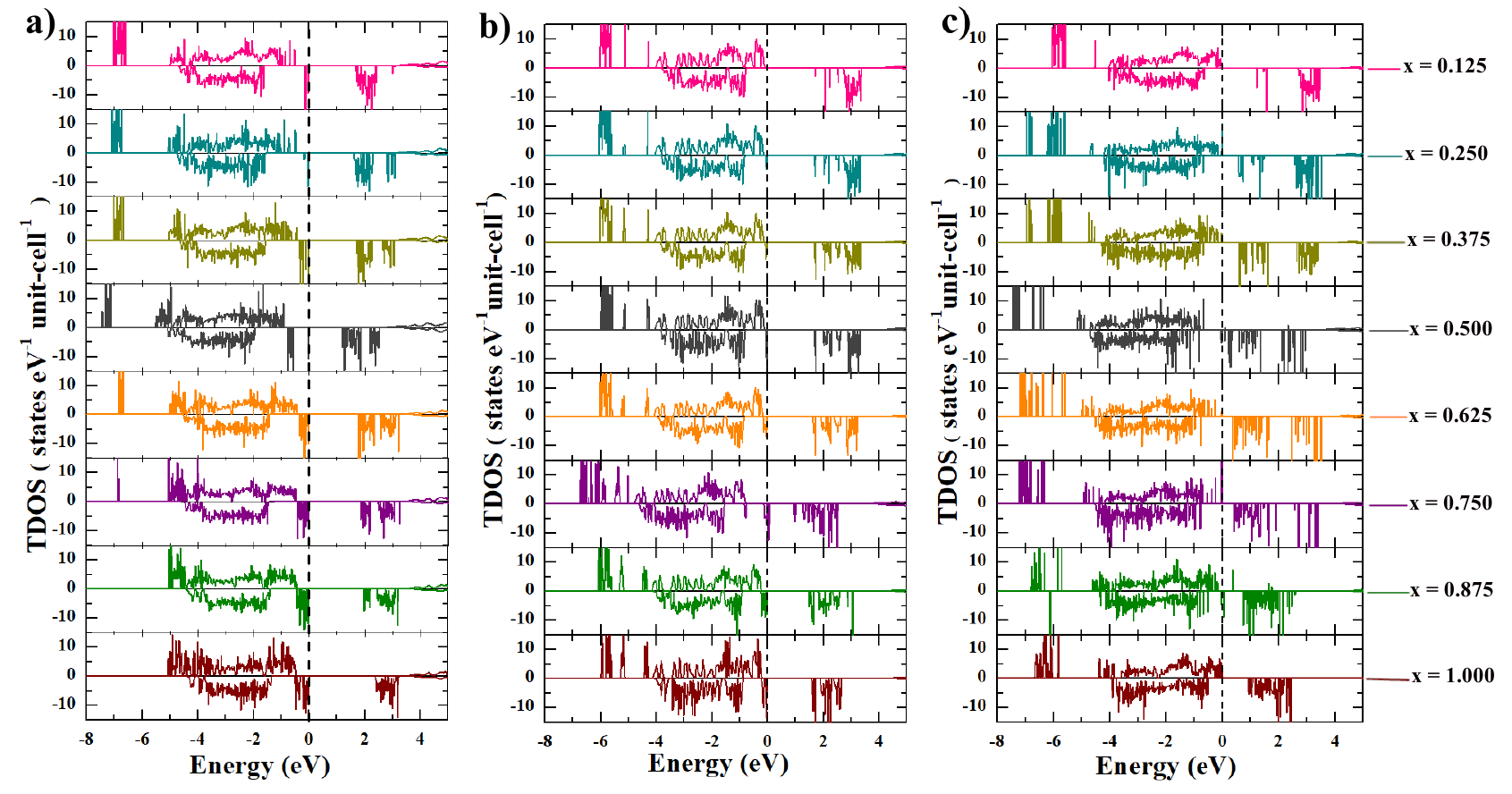}
\caption{The calculated total density of states for Li$_{y}$Fe$_{(1-x)}$Co$_{x}$O$_4$ systems where (a) {\it y = 5+x}, (b) {\it y = 5} and (c) {\it y = 5-x} ; $x = 0.00$, 0.125, 0.250, 0.375, 0.500, 0.625, 0.750, 0.875, 1.00 obtained from spin-polarized GGA + {\it U } calculation.  }
\label{fig:3}
\end{figure*}

 In the present study, we have considered  Li$_y$Fe$_{1-x}$Co$_x$O$_4$ with varying Co/Li concentrations. It is crucial to investigate the changes in the electronic structure of the materials by varying the concentration of Li and Co because they play vital roles in deciding the electrochemical properties of the cathode, especially the electrochemical potential. Fig.~\ref{fig:3} contains the total density of states (TDOS) for all the compounds considered in this work. Since a lower value of bandgap will improve electronic conductivity, which is an essential property of potential cathodes, the bandgap values from TDOS of all the systems are reported in Table~\ref{tab:3}. The variation in the bandgap values of these materials does not follow any systematic trend with Co concentration. It has to be noted that, along with Co, Fe and Li concentrations also change from one system to another and the cumulative effect comes into the picture to decide the bandgap values. Altogether, one can notice that, while increasing Co concentration (from top to bottom on each panel in Fig.~\ref{fig:3}) in Li$_{(5+x)}$Fe$_{(1-x)}$Co$_{x}$O$_4$ (Fig.~\ref{fig:3}(a)), Li$_{5}$Fe$_{(1-x)}$Co$_{x}$O$_4$ (Fig.~\ref{fig:3}(b)), and Li$_{(5-x)}$Fe$_{(1-x)}$Co$_{x}$O$_4$ (Fig.~\ref{fig:3}(c)), the bandgap mostly reduced to a minimum value and increase again with Co substitution (this happens twice in Li$_{(5-x)}$Fe$_{(1-x)}$Co$_{x}$O$_4$ system i.e., it have two minima one at $x = 0.5$ and the another at $x=0.875$ ). If we increase the Li concentration (from Fig.~\ref{fig:3}(a) to (c)) for a particular Co concentration, the bandgap gets reduced in most of the systems (exception: Li$_{5.25}$Fe$_{0.75}$Co$_{0.25}$O$_4$ and Li$_{5.5}$Fe$_{0.5}$Co$_{0.5}$O$_4$). The key takeaway information from the TDOS calculation is that the partial substitution of Co at  Fe sites reduces the bandgap value and hence one would expect better electronic conductivity than that of Li$_5$FeO$_4$ and fully Co substituted systems. For further analysis, we have selected the parent system Li$_5$FeO$_4$ along with smaller bandgap compositions such as Li$_{5.5}$Fe$_{0.5}$Co$_{0.5}$O$_4$, Li$_{5}$Fe$_{0.25}$Co$_{0.75}$O$_4$, and Li$_{4.5}$Fe$_{0.5}$Co$_{0.5}$O$_4$. Apart from the changes in the DOS due to an increase in the number of $d$ electrons by Co substitution and the variation of $s$ electron density by varying the Li concentration also reflected in the DOS curves and the major change in the TDOS plot for the selected Co substituted Li$_5$FeO$_4$ is the reduction in the bandgap value over the parent compound.   
 
 \par
\begin{figure*}[h]
\centering
\includegraphics[scale=1]{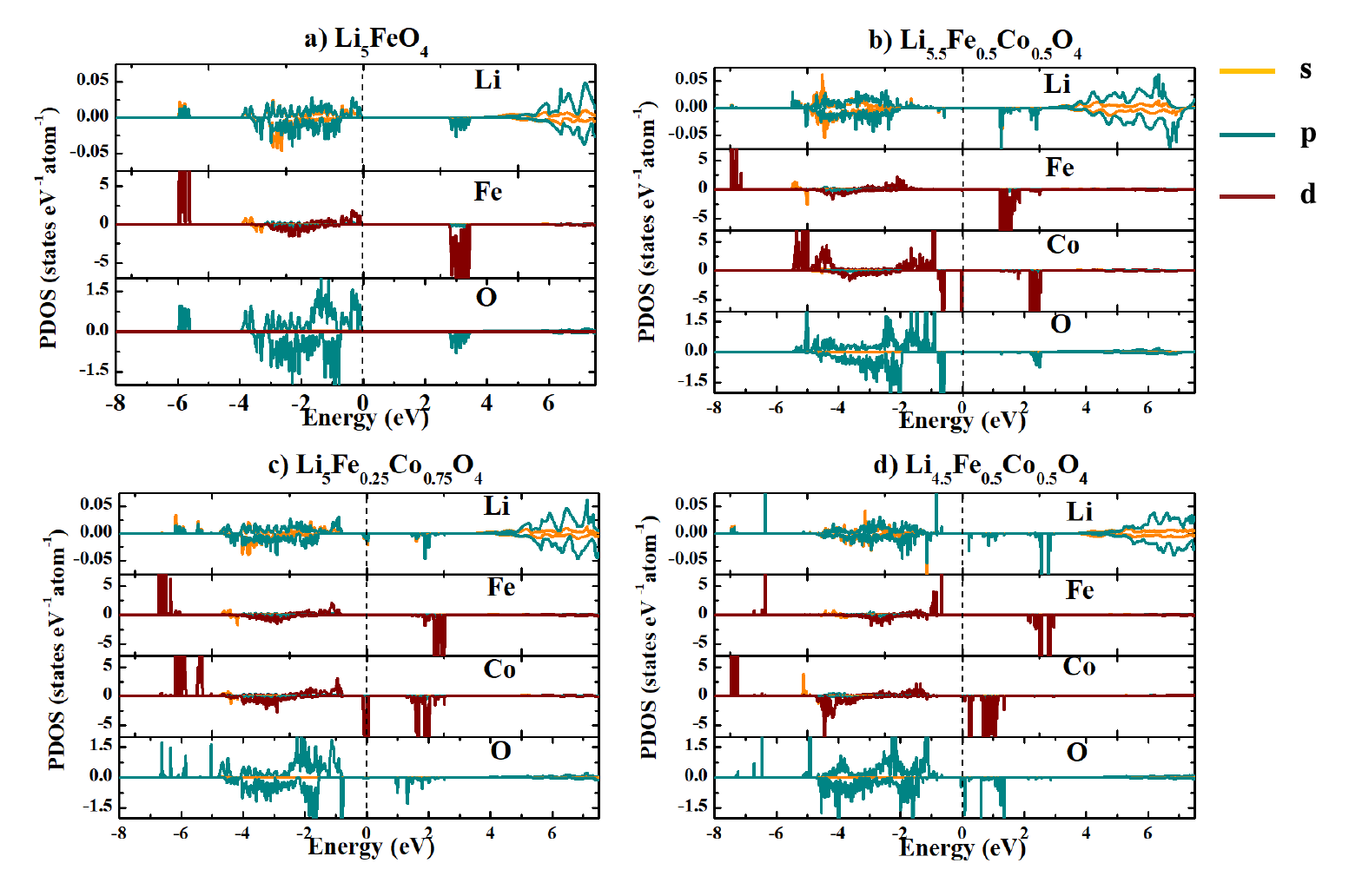}
\caption{The site, orbital and spin projected partial density of states for (a)  Li$_5$FeO$_4$,  (b) Li$_{5.5}$Fe$_{0.5}$Co$_{0.5}$O$_4$, (c) Li$_{5}$Fe$_{0.25}$Co$_{0.75}$O$_4$ (c), and (d)  Li$_{4.5}$Fe$_{0.5}$Co$_{0.5}$O$_4$ obtained from GGA+$U$ method with ferromagnetic configuration.}
\label{fig:4}
\end{figure*}

 In order to understand the relative contribution of constituent atoms on TDOS and gain more insight into the variation of electronic structure with Li and Co concentration change in Li$_5$FeO$_4$,  we have calculated the spin, site, and orbital projected density of states (PDOS) for the selected systems and depicted in Fig.~\ref{fig:4}. The corresponding PDOS for Li$_5$FeO$_4$ show that the valence band maximum (VBM) and conduction band minimum (CBM) are dominantly contributed by Fe-{\it 3d} and O-{\it 2p} states. The electronic states from Fe-{\it 3d} and O-{\it 2p} orbitals have energetically degenerated in the VB in the energy range 0.0 to $-$ 3.9 eV and $-$ 5.6 to $-$ 5.9 eV. These degenerated electronic states in the VB region indicate the presence of covalent bonding interaction between Fe and O in this system. The contribution from Li-$s$ states in the VB is relatively minimal, and this indicates the possibility of donation of Li valence electrons to neighboring O atoms, forming ionic bonding between them. As shown in Fig.\ref{fig:4}(b), the  PDOS of Li$_{5.5}$Fe$_{0.5}$Co$_{0.5}$O$_4$, the minority spin states of Co-{\it 3d} electrons fall on the VBM. It may be noted that the PDOS contribution from Co-{\it 3d} electrons are more than those from  Fe-{\it 3d} electrons for both the spin channel in the VB of Li$_{5.5}$Fe$_{0.5}$Co$_{0.5}$O$_4$. This is expected because there is one extra electron present in Co ($ Z = 27$) than that in Fe ($Z = 26$). Moreover, as Fe is usually in a +3 oxidation state (see section~\ref{Oxidation states analysis}) in this system,  we have modeled Li$_{5.5}$Fe$_{0.5}$Co$_{0.5}$O$_4$ in such a way that the Co is in +2 oxidation state. This will also result in more occupied Co-{\it 3d} states in this material. In other words, more density of states contribution from Co-{\it 3d} electrons in the VB suggests that the Co will be in +2 oxidation state in this system and is consistent with the conclusion arrived from charge analysis in section~ \ref{Oxidation states analysis}.   From the PDOS of Li$_{5.5}$Fe$_{0.5}$Co$_{0.5}$O$_4$ shown in Fig.\ref{fig:4}(c), we found that  the Co-{\it 3d} and O-{\it 2p} states are energetically degenerate in the energy range $-$0.45 to $- $5.46 eV while the Fe-{\it 3d} and O-{2p} states are in the range of $-$1.4 to $-$ 4.7 eV. As in  Li$_5$FeO$_4$, these energetically degenerate states from  Fe/Co and O, along with the spatial presence of Fe/Co and O adjacent to each other, indicate the presence of covalent bonding between Fe/Co and O and that is mainly responsible for stabilizing these compounds. On the other hand, the small value of density of states present at the VB of  Li site over its neutral state indicates an ionic bonding between Li and O, which is consistent with the electronegativity difference between these atoms. When we look into the PDOS of Li$_{5}$Fe$_{0.25}$Co$_{0.75}$O$_4$ in Fig.\ref{fig:4}(c), similar to  Li$_{5.5}$Fe$_{0.5}$Co$_{0.5}$O$_4$, the minority spin Co-{\it 3d} electron states fall at  the VBM. The value of PDOS from the  Co-{\it 3d} electrons  in the VB gets decreased in Li$_{5}$Fe$_{0.25}$Co$_{0.75}$O$_4$ compared to that in  Li$_{5.5}$Fe$_{0.5}$Co$_{0.5}$O$_4$ as shown in Fig.\ref{fig:4}. This is an indication for enhancement in electron transfer from Co site to the host lattice in Li$_{5}$Fe$_{0.25}$Co$_{0.75}$O$_4$ as we expected. Besides, the transfer of {\it 3d}-electrons from both Fe and Co sites to O sites and the degenerate nature of electronic states from Fe/Co-{\it 3d} with O-{\it 2p} states in the entire VB suggest the presence of an iono-covalent bond between Fe/Co and O. We have also plotted the PDOS for Li$_{4.5}$Fe$_{0.5}$Co$_{0.5}$O$_4$ in Fig.~\ref{fig:4}(d). In this system, the electronic states from Li, Co, and O are distributed in the entire valence band region. It is expected that the number of occupied states from Co-{\it 3d} and  Fe-{\it 3d} electrons are almost the same due to the presence of an equal number of valence electrons at both sites even though the oxidation states are different ( Fe (+3)  and Co (+4) site). As identified for other systems, the bonding nature of Co/Fe with O is iono-covalent in Li$_{4.5}$Fe$_{0.5}$Co$_{0.5}$O$_4$. However, as the Co is in +4 oxidation state,  the ionicity of Co$-$O interactions increases compared to other systems considered in the present study.

 \par
 On a broader view, in the Co substituted systems, the Co-{\it 3d} states fall on the VBM while for the pristine Li$_5$FeO$_4$, Fe and O states fall on the VBM. The chemical bonding analysis based on PDOS suggests that the interaction between Fe/Co and O is iono-covalent. The existence of covalency can be deduced from the presence of  O-{\it 2p} and Fe/Co-{\it 3d} states in a degenerate manner in the entire valence band in Fig.~\ref{fig:4}.  The bonding interaction between Li and O shows an ionic character, as evidenced by the limited number of states in the VB of the PDOS from Li. Moreover, in  Li$_5$FeO$_4$, the Fe-{\it 3d} states in the valence band are distributed unequally in the majority and minority spin states and giving a net magnetic moment of 4.796 $\mu_B$.  Similar to  Li$_5$FeO$_4$, all the Co substituted derivatives are also possess a net magnetic moment due to the exchange splitting of majority spin and minority spin electrons in the Fe/Co sites.

\par
 The electronic properties of cathode materials have an important role in their battery applications, including charge transport, especially in the case of predicting the possibility of anionic redox in Li-rich cathodes. The presence of  2{\it p}-states of O along with the {\it 3d}-states of Fe/Co metal at the topmost region of the valence band elucidates the possibility of both anionic and cationic redox in these materials and such behavior is visible in Li$_{4.5}$Fe$_{0.5}$Co$_{0.5}$O$_4$.  Whereas, in Li$_{5.5}$Fe$_{0.5}$Co$_{0.5}$O$_4$ and Li$_{5}$Fe$_{0.25}$Co$_{0.75}$O$_4$ as shown in Fig.~\ref{fig:4} (b) and (c), the  O-{\it 2p}, states are falling just below the narrow {\it 3d} states of Co at the VBM. So these materials are also capable of exhibiting anionic redox at higher redox potential~\cite{sun2021decoupling}. More studies are needed to confirm the anionic redox during the delithiation process, and it will be discussed in detail in the following sections.
 
\begin{table*}[]
\centering
\caption{The calculated values of bandgap (eV) for  Li$_{y}$Fe$_{(1-x)}$Co$_{x}$O$_4$ systems (where {\it y = 5+x},  {\it y = 5} and {\it y = 5-x} $5-x$; $x = 0.00$, 0.125, 0.250, 0.375, 0.500, 0.625, 0.750, 0.875, 1.00) estimated from total density of states.}
\begin{threeparttable}
\label{tab:3}
\begin{tabularx}{409pt}{ l c c c c c c c c c }
\hline
\multirow{2}{*}{Compound} & \multicolumn{9}{c}{  bandgap (eV) }   \\\cline{2-10}

  & x =  0.00 & 0.125 & 0.250 & 0.375 & 0.500 & 0.625 &  0.750 & 0.875 & 1.00 \\ 
\hline
Li$_{(5+x)}$Fe$_{(1-x)}$Co$_{x}$O$_4$  &  & 1.70 & 1.64 & 1.71  & 1.22 & 1.81 & 1.86 & 1.99 & 2.40 \\
Li$_{5}$Fe$_{(1-x)}$Co$_{x}$O$_4$  & 2.62 & 1.72 & 1.68 & 1.69 & 1.69 & 1.64 & 0.01 & 1.50 & 1.57\\
Li$_{(5-x)}$Fe$_{(1-x)}$Co$_{x}$O$_4$ &  & 1.23 & 0.60 & 0.63 & 0.07 & 0.37 & 0.42 & 0.07 & 0.92\\
\hline
\end{tabularx}
\label{tab:3}
\end{threeparttable}
\end{table*}

\subsubsection{Oxidation states analysis  of Fe and Co ions in Li$_{y}$Fe$_{(1-x)}$Co$_{x}$O$_4$}
\label{Oxidation states analysis}

\par
In the present study, we have used various charge analysis schemes such as  Born effective charge (BEC), Bader effective charge (BC), Mulliken charge (MC), and L{\"o}wdin charge (LC) population analyses~\cite{king1993theory,bader1985atoms,henkelman2006fast} to investigate the charge states of Co, Fe, and O ions for selected compositions. The BEC is calculated using the Berry phase approach of the “Modern theory of polarization” ~\cite{king1993theory} and are listed in Table~\ref{tab:4}. We have selected Li$_5$FeO$_4$ and Li$_{5.5}$Fe$_{0.5}$Co$_{0.5}$O$_4$ systems for the BEC calculations due to their finite bandgap values (see Table~\ref{tab:3}). From the diagonal component of BEC in Table~\ref{tab:4}, one can notice that Li has a value of around 0.9\,e, which is close to the formal oxidation state of Li$^{1+}$ in these materials. Similarly, O also shows a BEC value of around $-$1.8\,e, which is also close to its formal oxidation state of $-2$.  It indicates that the Li$-$O bond shows a prominent ionic character. For the pure ionic case, the diagonal components are equal, and the off-diagonal elements are expected to be zero. But, the presence of a non-equivalent diagonal component of BEC at both Li and O sites along with a finite off-diagonal component of BEC in both sites indicates that finite covalency is also present in the Li$-$O bond. The Fe atom has a BEC value of 2.3\,e in both the materials suggesting that Fe is in a +3 oxidation state. The reduction in the BEC value compared with the formal oxidation state of Fe indicates an iono-covalent bond between Fe with its surrounding O and this is discussed in more detail in the following section (see section \ref{Analysis of chemical bonding}).  It is interesting to note that the average value of the diagonal component of BEC at the Co site is around 1.7\,e, which indicates that the Co will be in a +2 oxidation state in Li$_{5.5}$Fe$_{0.5}$Co$_{0.5}$O$_4$. The slight difference between the calculated BEC and expected oxidation state (+2) comes from Co$-$O covalent interaction.\\
\
 In the BC analysis, the continuous electron density is partitioned and assigned to individual atoms by defining regions bounded by zero flux surfaces of charge density~\cite{bader1985atoms,aboud2010density} called basins and integrated charge density within these basins are assigned to each atom. The calculated BC, MC, and LC values at the Fe, Co, and O sites in selected systems are tabulated in Table~\ref{tab:5}. From Table~\ref{tab:5}, it can be noticed that the charge at the Fe sites is almost the same in all the selected systems as expected, and hence the oxidation state of Fe is assigned to be +3 as in Li$_5$FeO$_4$ due to its consistent atomic environment. The charge on O is also found to be varying with the Li concentration. This is because the O ions are directly connected to the Li-ions and also indicate the participation of O in the redox reaction. The Li and O interaction is ionic in nature, as discussed in the following sections. So the changes in the Li concentration will directly affect the charge of O, apart from the apparent change of oxidation state of Fe/Co. Interestingly, the Co atoms which do not have any direct bond with O are also affected by the Li concentration change.  A comparative analysis of charge at Co sites indicates that an inverse relation exists between the charge at the Co sites with the concentration of  Li. It is a direct indication of the decrease in oxidation number of Co atoms when more Li is taking part in the electron transfer to the O. I.e., the Co will be in +2, +3, +4 oxidation states, respectively in Li$_{5.5}$Fe$_{0.5}$Co$_{0.5}$O$_4$, Li$_{5}$Fe$_{0.25}$Co$_{0.75}$O$_4$  and  Li$_{4.5}$Fe$_{0.5}$Co$_{0.5}$O$_4$ systems. This verifies our methodology of tuning the oxidation states of the Co by controlling the number of adjacent Li-ions. One can find that the neighboring Co ion tries to neutralize the system during lithiation/delithiation since the Co can easily be oxidized compared to the O atoms. But there is a finite charge difference at the O site for different Li concentrations. This may be because the Li$-$O bond is ionic, whereas the Co$-$O bond is iono-covalent, as we discuss in detail in the following section~\ref{Analysis of chemical bonding}. So the charge will not be accumulated on the O site as in a Li$-$O bond while forming a Co$-$O bond.   This may also be the reason behind the charge on the Co/Fe being less than that of the classical oxidation states of the Co/Fe. In all the considered systems, the atomic charges calculated using all three methods exhibit a consistent trend, i.e., the charge at the Fe ion is almost the same in all the systems, and similarly,  the charge at the Co site increase and the negative charge on O decreases with a decrease in Li concentration. But the numerical values of the charges estimated from different methods are slightly different.

\begin{table*}[]
\centering
\caption{Calculated Born effective charge (in e) for  Li$_5$FeO$_4$ and  Li$_{5.5}$Fe$_{0.5}$Co$_{0.5}$O$_4$  systems }
\begin{threeparttable}
\label{tab:4}
\begin{tabularx}{500pt}{ l c c c c c c c c c c c }
\hline
\multirow{2}{*}{Compound} & \multirow{2}{*} {atom} & \multicolumn{10}{c}{ Born effective charge $Z^\ast$ (e)}   \\\cline{3-12}

 & & average & Z$_{xx}$ & Z$_{yy}$ & Z$_{zz}$ & Z$_{xy}$ & Z$_{xz}$ &  Z$_{yx}$ & Z$_{yz}$ & Z$_{zx}$ & Z$_{zy}$ \\ 
\hline
Li$_5$FeO$_4$ & Li  & 0.960 & 1.055 & 0.868 & 0.956 &  0.167 &  -0.055 &  -0.032 &  -0.119 &  0.053  & -0.218 \\
&Fe & 2.263 &2.458 &  2.144 & 2.188 &- 0.119 &- 0.016 & -0.274 & 0.042 &  - 0.100 &  -0.094\\
& O  & -1.820 & -1.888 & -1.849 & -1.724 & -0.212 & -0.384 & -0.305 &  -0.288 & -0.323 &  -0.434 \\
Li$_{5.5}$Fe$_{0.5}$Co$_{0.5}$O$_4$ & Li  & 0.941 &  0.945 &  1.030 & 0.847 & 0.042 & -0.166 & 0.071 & -0.068 &  -0.073 & 0.101  \\
&Fe & 2.323  & 2.503 & 2.315 & 2.150 & 0.140 & -0.0815 & 0.265 & -0.095 &  -0.166 & 0.016   \\
& Co & 1.703  &  1.902 &  1.695 & 1.511 & -0.178 &  -0.153 & -0.265 & 0.056 & -0.132 & 0.073 \\
& O  & -1.814  &  -1.847 & -1.641 &  -1.953 & -0.259 &  0.206 & -0.235 & 0.267 & 0.341 & 0.251\\
\hline
\end{tabularx}
\label{tab:4}
\end{threeparttable}
\end{table*}

\begin{table*}[]
\centering
\caption{The calculated  charge (in e)  at  Fe, Co and O sites in  Li$_{y}$Fe$_{(1-x)}$Co$_{x}$O$_4$ systems obtained  from Bader effective charge, Mulliken charge, and L{\"o}wdin  charge analysis scheme.}
\begin{threeparttable}
\label{tab:5}
\begin{tabularx}{355pt}{ c c c c c c c c c c}
\hline
\multirow{2}{*}{compound} & \multicolumn{3}{c}{Bader charges (e)} & \multicolumn{3}{c}{Mulliken charges (e)}  & \multicolumn{3}{c}{L{\"o}wdin charges (e)}\\\cline{2-10}
&   Fe   & Co & O & Fe & Co & O & Fe & Co & O  \\ 
\hline
Li$_5$FeO$_4$  & 1.50 & & -1.46 & 1.75 & & -1.16 & 1.61 & & -1.18  \\
Li$_{5.5}$Fe$_{0.5}$Co$_{0.5}$O$_4$  & 1.50 & 1.01 & -1.51 & 1.78 & 1.39 & -1.42 & 1.62 & 1.20 & -1.35 \\
Li$_{5}$Fe$_{0.25}$Co$_{0.75}$O$_4$  & 1.49 & 1.29 & -1.43 & 1.74 & 1.63 & -1.36 & 1.61 & 1.48 & -1.28 \\
Li$_{4.5}$Fe$_{0.5}$Co$_{0.5}$O$_4$  & 1.50 & 1.46 & -1.35 & 1.74 & 1.74 &  -1.29 & 1.61 & 1.60 & -1.22 \\ 
\hline
\end{tabularx}
\label{tab:5}
\end{threeparttable}
\end{table*}

\subsubsection{Analysis of chemical bonding  in  Li$_{y}$Fe$_{(1-x)}$Co$_{x}$O$_4$ system}
\label{Analysis of chemical bonding}

\par
\begin{figure*}[h]
\centering
\includegraphics[scale=1]{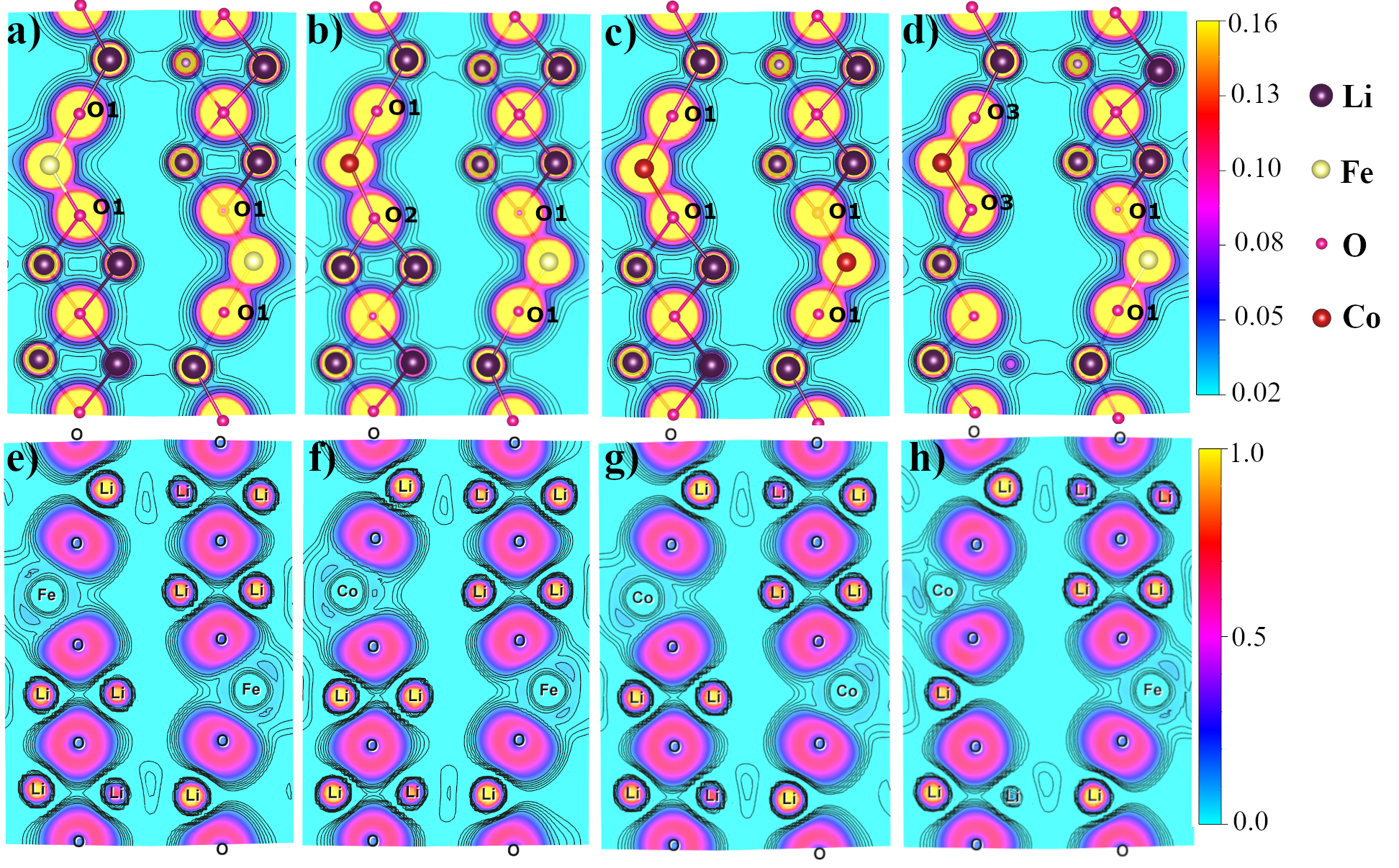}
\caption{The contour map of   (a-d)  charge density in e/{\AA}\textsuperscript{3} and (e-h) electron localization functions for selected systems ( (a and e) Li$_5$FeO$_4$, (b and f) Li$_{5.5}$Fe$_{0.5}$Co$_{0.5}$O$_4$, (c and g) Li$_{5}$Fe$_{0.25}$Co$_{0.75}$O$_4$, (d and h ) Li$_{4.5}$Fe$_{0.5}$Co$_{0.5}$O$_4$ ) in a plane where we can see all the distinct atom.}
\label{fig:5}
\end{figure*}

 
The bonding interaction between the transition metals and oxygen in transition metal oxide cathode materials would influence the structural stability as well as the electrochemical potential during battery operation. Moreover, the bonding interaction of Li with its surrounding oxygen atoms plays a vital role in the lithiation/delithiation process of LIB. We have carried out detailed chemical bonding and bond strength analysis between constituents in Li$_5$FeO$_4$ and its Co substituted derivatives as a function of Li concentration using different techniques in this section. The charge density analysis enables one to get a detailed view of the bonding interaction between the constituents in a material. In Fig:\ref{fig:5}, we have given the charge density plot for a plane where one can see the chemical bonding interaction  between the constitutions in  (a) Li$_5$FeO$_4$, (b) Li$_{5.5}$Fe$_{0.5}$Co$_{0.5}$O$_4$, (c) Li$_{5}$Fe$_{0.25}$Co$_{0.75}$O$_4$, and (d) Li$_{4.5}$Fe$_{0.5}$Co$_{0.5}$O$_4$. The planes are selected in such a way that the Fe and Co come in the same plane and also, the Fe/Co$-$O and Li$-$O bonds are visible. From the charge density plot, one can notice that the charges are spherically distributed around the Li site in all the considered systems. There is no finite charge distribution exists between the Li and oxygen and this is a clear indication of the presence of ionic interactions between Li$-$O  in all the considered systems as expected based on the electronegativity difference between  Li and O. Whereas, the bonding interactions between Fe/Co and O show prominent covalent character. It can be identified from the finite anisotropic charge density distributions between these atoms irrespective of the Li/Fe/Co concentration. A comparative analysis of charge density plots among the selected systems indicates that the charge density distribution between Fe and O does not change by Li and/or Co concentration change in Li$_{y}$Fe$_{(1-x)}$Co$_{x}$O$_4$. At the same time, the covalency of the Co$-$O bond gets weaker/stronger during the addition/removal of Li. This is because the Li addition/removal changes the oxidation state of the Co.

\par
 If one substitutes Li in the vacant sites available in Li$_{y}$Fe$_{(1-x)}$Co$_{x}$O$_4$, the coordination of some of the O gets changed from 6 (coordinated by 5 Li and 1 Co atoms) to 7 (coordinated by 6 Li and 1 Co atoms). We call this six atom coordinated O as O1 and seven atoms coordinated O as O2. In contrast, the removal of Li leads to a coordination change for some of the oxygen from 6  to 5 (coordinated by 4 Li and 1 Co atoms), and we call these five atoms coordinated O as O3.  In  Li$_{5.5}$Fe$_{0.5}$Co$_{0.5}$O$_4$  (Fig:\ref{fig:5}(b)),  within the projected plane, one of the O is bonded to Co with coordination number 7. This Co$-$O2 bond shows a depletion of charge density between atoms compared with the Co$-$O1 bond as evident from (Fig:\ref{fig:5}(b) and thereby the reduced covalency. This reduction in covalency in Co$-$O2 bond over Co$-$O1 bond is due to a comparatively higher negative charge on the O2 site from an extra Li$-$O ionic interaction. The other O with a coordination number of 6 (O1) does not show any change in charge density between the Co$-$O1 bonds during lithiation. Similarly, in Li$_{4.5}$Fe$_{0.5}$Co$_{0.5}$O$_4$ (Fig:\ref{fig:5}(d)), the two O3  bonded with Co make Co$-$O3 covalent bond. This bond is relatively more covalent than the Co$-$O1 bond due to a small negative charge at the O site resulting from the absence of a Li$-$O bond in comparison with that in  Li$_{5}$Fe$_{0.25}$Co$_{0.75}$O$_4$ (Fig: \ref{fig:5}(c)) system.  One may also note that two Li are missing in Li$_{4.5}$Fe$_{0.5}$Co$_{0.5}$O$_4$ (Fig:\ref{fig:5}(d)) in the selected plane compared to other systems in Fig: \ref{fig:5}. Among them, one has been removed, and the other one has been moved away from the given plane after structural relaxation.

\par
In order to get further insights into the bonding mechanism, we have plotted the electron localization function (ELF) for the selected systems in  Fig:\ref{fig:5}(e)$-$(h), the same planes as that used for charge density analysis. In a chemical system, the ELF value helps to distinguish between homogeneously distributed electrons  (ELF = 0.5) and fully localized electrons (ELF = 1). Every system considered in the present study shows a maximum ELF value at the Li sites and this is due to the well-localized 1{\it s} electrons (included along with 2{\it s} electrons in the present study). At the same time, one cannot find any localized electron pairs in the region between Li and O atoms indicating the ionic character of Li$-$O bonds. The ELF value greater than 0.5 is found at the O site indicating the presence of localized 2$p$ electrons originating mainly by the transfer of electrons from cation sites to O, making a closed-shell configuration. This once again confirms the ionic bonding interaction between Li and O. In contrast, there are non-localized electron distributions around the Fe/Co sites, indicated by the small value for the ELF distribution at this site. Also, there are finite ELF distributions between Fe/Co and O atoms. The anisotropic ELF distribution between Fe/Co and O, along with the noticeable value of ELF between Fe/Co and O, suggests the presence of some covalent interactions between Fe/Co and O atoms. 

\begin{figure*}[h]
\centering
\includegraphics[scale=1]{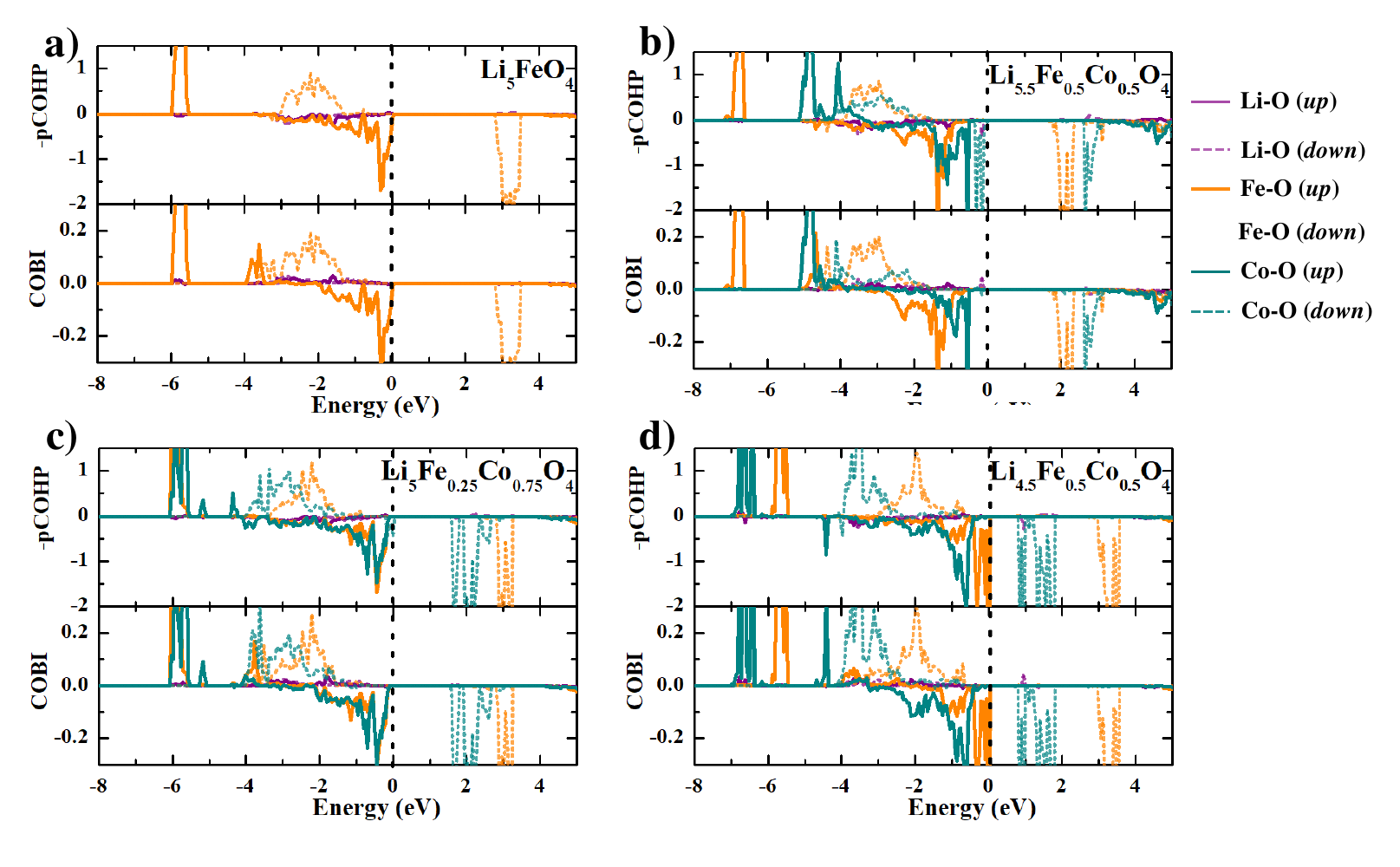}
\caption{The calculated  projected crystal orbital Hamilton population (pCOHP)  and crystal orbital bond index (COBI) between constituents in (a) Li$_5$FeO$_4$ ,  (b) Li$_{5.5}$Fe$_{0.5}$Co$_{0.5}$O$_4$,  (c) Li$_{5}$Fe$_{0.25}$Co$_{0.75}$O$_4$, and  (d) Li$_{4.5}$Fe$_{0.5}$Co$_{0.5}$O$_4$ using LOBSTER package with input from spin polarised GGA+ {\it U } calculation with complete structural optimization. }
\label{fig:6}
\end{figure*}

Crystal orbital bond index (COBI) is a new method in solid-state physics for quantifying covalent bonding in materials~\cite{muller2021crystal} and is implemented in the LOBSTER package~\cite{dronskowski1993crystal}. It is a generalization and extension of the concept of the bond index by Wiberg and Mayer to the case of crystalline solids~\cite{muller2021crystal}. The COBI for selected systems is given in Fig.~\ref{fig:6}. Furthermore, in order to quantify the covalent interactions in the considered systems, the integrated value of COBI below the Fermi level is given in Table~\ref{tab:6}. A high ICOBI value for a bonding pair indicates covalent interaction, whereas a smaller ICOBI value indicates ionic interactions. The metallic bonding interaction between bonding pairs is represented by an ICOBI value that is smaller than that of covalent interaction~\cite{simons2021bonding}. In Li$_5$FeO$_4$, the Li$-$O has a relatively small ICOBI value and therefore, the Li$-$O interactions will be primarily ionic, as expected. However, the ICOBI value for the Fe$-$O bonding pair is in-between that of pure covalent and ionic bonds. Hence, one can conclude that the bonding interaction between Fe and O is of iono-covalent nature. It may be noted that, this is consistent with the conclusion arrived from our DOS analysis and various charge analyses made in the previous section. Also, from the calculated value of ICOBI for Fe$-$O and Co$-$O bonding pairs in Li$_{5.5}$Fe$_{0.5}$Co$_{0.5}$O$_4$, we found that the Co$-$O bond is less covalent than the Fe$-$O bond, albeit it is more covalent in Li$_{5}$Fe$_{0.25}$Co$_{0.75}$O$_4$ and Li$_{4.5}$Fe$_{0.5}$Co$_{0.5}$O$_4$. 
\par
The previous studies on LIB cathodes indicate that the electrochemical potential of a system is related to the degree of iono-covalent nature of TM$-$O bonds. A more ionic nature of TM$-$O interaction will have a higher potential~\cite{grimaud2016anionic,goodenough2010challenges}, whereas the strong covalent nature of the interaction between TM$-$O will promote anionic redox in cathode materials~\cite{grimaud2016anionic}. So, our COBI analysis of  bonding pairs in  Li$_{y}$Fe$_{(1-x)}$Co$_{x}$O$_4$ suggests that,  among the four selected systems, Li$_{5}$Fe$_{0.25}$Co$_{0.75}$O$_4$ and Li$_{4.5}$Fe$_{0.5}$Co$_{0.5}$O$_4$ will be more susceptible to have anionic redox during lithiation/delithiation process. 

\begin{table}[]
\centering
\caption{The calculated integrated crystal orbital Hamilton population (ICOHP in eV/bond)  and integrated  crystal orbital bond index (ICOBI)  for the bonding pairs in Li$_5$FeO$_4$, Li$_{5.5}$Fe$_{0.5}$Co$_{0.5}$O$_4$, Li$_{5}$Fe$_{0.25}$Co$_{0.75}$O$_4$, and Li$_{4.5}$Fe$_{0.5}$Co$_{0.5}$O$_4$.}
\begin{threeparttable}
\label{tab:6}
\begin{tabularx}{230pt}{ c c c c c c c c }
\hline
Compound & Interaction & -pICOHP & -ICOBI \\
\hline
Li$_5$FeO$_4$  & Li$-$O &  0.946 &  0.124 \\
& Fe$-$O &  2.311 &  0.328 \\
Li$_{5.5}$Fe$_{0.5}$Co$_{0.5}$O$_4$  & Li$-$O &  0.936 & 0.121 \\
& Fe$-$O & 2.381 & 0.337 \\
& Co$-$O & 1.694 & 0.229 \\
Li$_{5}$Fe$_{0.25}$Co$_{0.75}$O$_4$  & Li$-$O &  0.999  &  0.123\\
& Fe$-$O &  2.333 & 0.325 \\
& Co$-$O &  2.730 & 0.380 \\
Li$_{4.5}$Fe$_{0.5}$Co$_{0.5}$O$_4$ & Li$-$O & 1.143 & 0.137 \\
& Fe$-$O &  2.567 &  0.361 \\
& Co$-$O &  2.997 & 0.454  \\

\hline
\end{tabularx}
\label{tab:6}
\end{threeparttable}
\end{table}

\subsubsection{Bond strength analysis}
\label{Bond strength analysis}

\par
In order to have insight into the chemical bonding between the constituents in  Li$_{y}$Fe$_{(1-x)}$Co$_{x}$O$_4$ as a function of Li and Co concentration, we have calculated the average equilibrium bond length between constituents using spin polarised GGA+ {\it U } calculation with complete structural optimization and the calculated values for selected systems are given in Table~\ref{tab:7}. From this table, the equilibrium bond lengths between Fe and oxygen ions in the selected systems are almost the same (1.9 {\AA}). Moreover, the Co ions that possess a similar Li environment as that of Fe exhibit an average equilibrium bond length as same as that of the Fe$-$O bond. This indicates that the Li concentration plays a vital role in deciding the bond strength of Fe/Co$-$O bonds in these systems. It is worthy to note that the lowering of Li concentration decreases the Co$-$O bond length and hence it is expected to increase the bond strength. An increase in Li concentration leads to an increase in the Co$-$O bond length value as given in Table~\ref{tab:7} and hence expected a reduction in the strength of the Co$-$O bond and this is confirmed based on our crystal orbital Hamilton population (COHP) analysis described in the following section. A stronger TM$-$O bond can improve the structural stability of the cathodes during the delithiation process and this could explain why Co substitution leads to an increase in the capacity. The calculated bond length between Li and O is more than that of the Fe/Co$-$O bond and is almost the same in every system. A higher bond length between Li and its coordinating O weakens the bond strength and helps to delithiate easily.  More detailed studies on the bond strengths between constituents for selected systems in Li$_{y}$Fe$_{(1-x)}$Co$_{x}$O$_4$ have been presented in the following section on the basis of projected COHP (pCOHP) analysis. 
\begin{table}[]
\centering
\caption{The calculated equilibrium average bond lengths (in {\AA}) between constituents in Li$_{y}$Fe$_{(1-x)}$Co$_{x}$O$_4$  for  selected compositions obtained from spin-polarised GGA+ {\it U } calculation with complete structural optimization.}
\begin{threeparttable}
\label{tab:7}
\begin{tabularx}{200pt}{ c c c c}
\hline
 Compound  & Fe $-$ O  & Co$ -$O  &  Li$-$O \\ 
\hline
Li$_5$FeO$_4$  & 1.901 &  & 2.003  \\
Li$_{5.5}$Fe$_{0.5}$Co$_{0.5}$O$_4$  & 1.910 & 1.992 & 2.018  \\
Li$_{5}$Fe$_{0.25}$Co$_{0.75}$O$_4$  & 1.899 & 1.885 & 1.996 \\
Li$_{4.5}$Fe$_{0.5}$Co$_{0.5}$O$_4$  &  1.898 & 1.807 & 2.003  \\
\hline
\end{tabularx}
\label{tab:7}
\end{threeparttable}
\end{table}

In order to evaluate the bond strength of the bonding pairs and also to identify the bonding as well as antibonding/non-bonding states of the bonding pairs in Li$_{y}$Fe$_{(1-x)}$Co$_{x}$O$_{4}$, we have calculated the COHP. The COHP is the density of states weighted by the corresponding Hamiltonian matrix element and is indicative of the strength and nature of bonding. The bonding and antibonding interactions are represented by negative and positive values of the COHP, respectively, while non-bonding interactions are indicated by zero values of the COHP~\cite{dronskowski1993crystal}. The pCOHP enables us to analyze the chemical bonding in materials by re-extracting Hamilton-weighted populations from plane-wave electronic structure calculations~\cite{deringer2011crystal}. In the present work, we have calculated the energy-resolved pCOHP(E) between Li$-$O, Fe$-$O, and Co$-$O pairs for all the selected systems and plotted separately for spin-up and spin-down electrons in  Fig.~\ref{fig:6}. Here we have plotted $-$pCOHP; therefore, the positive value of the plot will be bonding states and the negative value of the plot will be antibonding states. If there is finite DOS in particular energy for a bonding pair and there is zero value of COHP in that energy in the pCOHP curve, this is the indication of non-bonding interactions.~\cite{steinberg2018crystal}. The system will attain maximum stability when all the bonding states are filled and the antibonding states are empty. In magnetic systems, in order to maximize the filling of bonding states, the energy levels are exchange splitted. Hence, we have presented the -pCOHP for majority spin electrons and minority spin electrons in the bonding phase pairs separately.  Among the bonding interactions between various bonding pairs in Li$_{y}$Fe$_{(1-x)}$Co$_{x}$O$_4$,  the Fe/Co$-$O bonding pair has strong bonding interaction and hence a large contribution of bonding states from Fe/Co$-$O bonding pair present in the VB region. Even though the bonding states are mainly dominated in the VB, a substantial antibonding character is also present in the VB, indicating that these systems will have reduced stability. This may be the reason why in the lithiation and delithiation process, these compounds also have stable structures by changing the oxidation states of transition metal ions. 

Moreover, the integrated COHP (ICOHP) scales with the bond strength. From the calculated values of ICOHP given in Table~\ref{tab:6}, one can notice that the Li$-$O interaction is weak compared to Fe/Co$-$O interactions in all the systems shown in Table~\ref{tab:6}. This indicates that these systems can delithiate without much structural collapse. The bonding interaction between Fe/Co with O plays an important role in deciding the stability of these compounds. When we analyze the ICOHP for the bonding pairs in the considered systems, we found that in Li$_{5.5}$Fe$_{0.5}$Co$_{0.5}$O$_4$, the Co$-$O interactions have a smaller ICOHP value than that of the Fe$-$O bonding pair and this suggests that the Co substitution will decrease the structural phase stability. However, when we reduce the Li concentration in the system, the ICOHP value for the Co$-$O bonding pair increases as in the case of Li$_{5}$Fe$_{0.25}$Co$_{0.75}$O$_4$ and Li$_{4.5}$Fe$_{0.5}$Co$_{0.5}$O$_4$ and hence Co$-$O bonding pair in these systems have higher ICOHP value than that of Fe$-$O bonding pairs. These suggest that during the delithiation process, the Co substitution improves the stability of these systems. This could explain why extra Li can be removed from Co substituted systems without destabilizing the structure. In other words, the structural stability will increase while increasing the charge on the Co atom via systematically decreasing Li concentration around it. This is as expected based on the bond length analysis described above.  

\subsection{Li-ion  diffusion in Li$_{y}$Fe$_{(1-x)}$Co$_{x}$O$_4$ }
\label{Li-ion  diffusion in Li$_{y}$Fe$_{(1-x)}$Co$_{x}$O$_4$}

\begin{figure*}[h]
\centering
\includegraphics[scale=1]{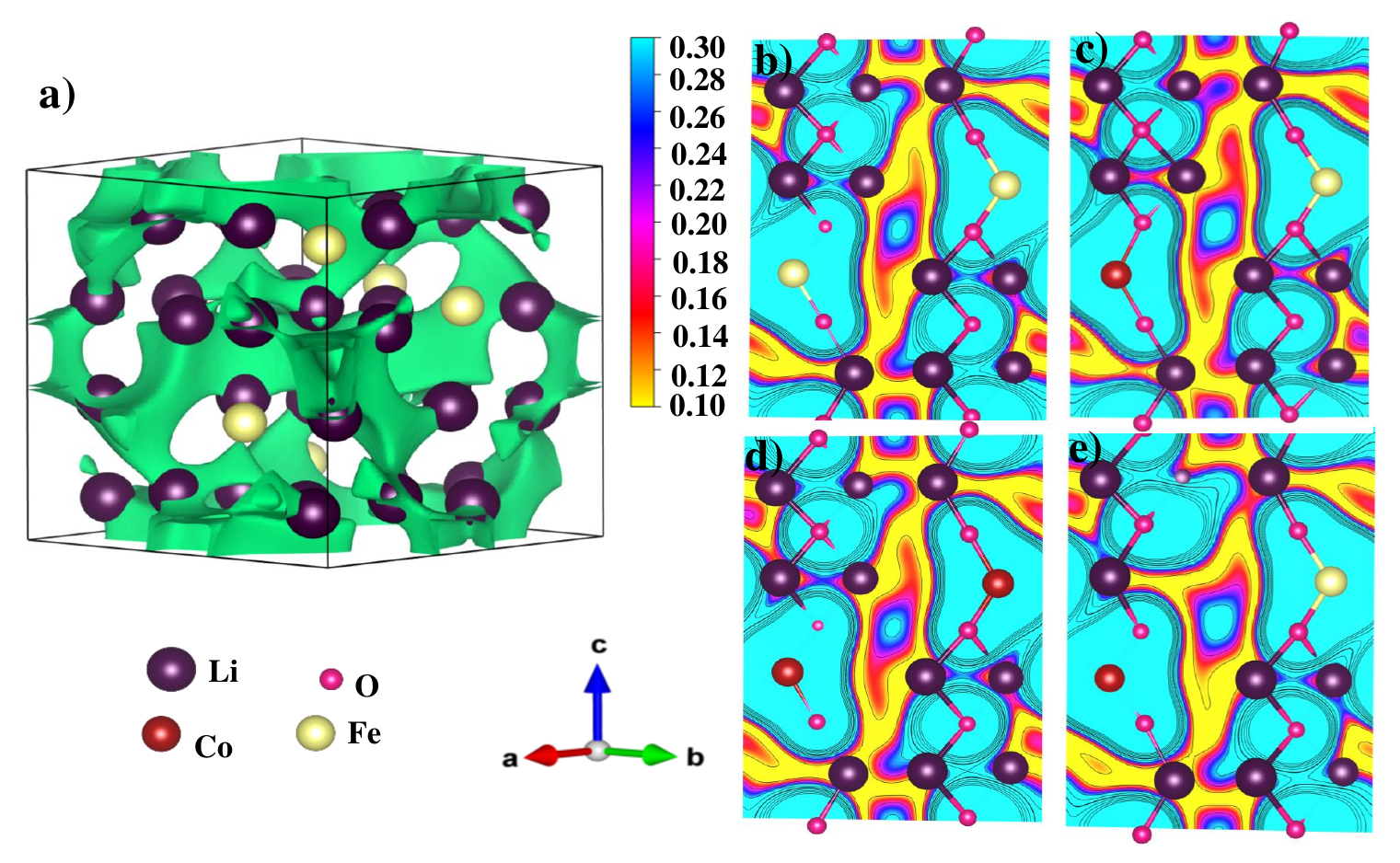}
\caption[c]{ (a) The crystal structure of Li$_5$FeO$_4 $ overlapped with its  isosurface of 3D bond valence sum (BVS) difference $\Delta|V|$.  The contour map of  BVS difference  $|\Delta V|$  for (b)  Li$_5$FeO$_4$,  (c) Li$_{5.5}$Fe$_{0.5}$Co$_{0.5}$O$_4$, (d)  Li$_{5}$Fe$_{0.25}$Co$_{0.75}$O$_4$, and (e)  Li$_{4.5}$Fe$_{0.5}$Co$_{0.5}$O$_4$  in selected planes}.
\label{fig:7}
\end{figure*}

\par

In order to gain knowledge about approximate Li diffusion pathways in the selected systems, we have calculated the three-dimensional spatial distribution of bond valence sum (BVS) using the program PyAbstantia~\cite{nishimura2017pyabstantia}. In this approach, one can obtain the deviation of BVS of a target atom from its ideal valency ($\Delta V$) as follows:

\begin{equation} \label{eu_eqn2}
|\Delta V| = |V_{Li}-V_{f}|,
\end{equation}

where $V_{Li}$ is the calculated BVS of Li in the system and $V_{f}$ is the formal valence state of Li. In a BVS map, the lower $|\Delta V|$  regions correspond to more stable sites of Li occupations and hence indicate Li-ion diffusion pathways~\cite{gamo2021effects,panigrahi2017sodium}. In Fig.~\ref{fig:7}(a), we have depicted  the  $|\Delta V|$ = 0.6 v.u. isosurface (green colored surfaces)  in the unit cell of Li$_5$FeO$_4 $  and it indicates the three-dimensional pathway of Li diffusion in Li$_5$FeO$_4$. Due to the electrostatic repulsion from  Fe/Co, the Li conduction along Fe/Co is limited. The vacancy ordering present in these materials eases the Li diffusion and thereby improves the ionic conductivity. In order to have a better understanding of the effect of substitution of Co with different valencies in Li migration, we have plotted the contour map of $|\Delta V|$ for a selected plane in the range of 0.1 v.u. to 0.3 v.u. in Fig.~\ref{fig:7}(b-e), where the region with lower $|\Delta V|$ (in yellow) indicates the lower-energy pathways for Li-ion migration.  Even though there is not much difference in the spatial distribution of bond valence sum in Fig.~ \ref{fig:7}(b$-$e), one can notice that the Li$_{5.5}$Fe$_{0.5}$Co$_{0.5}$O$_4$ ( Fig.~\ref{fig:7}(c)) and Li$_{4.5}$Fe$_{0.5}$Co$_{0.5}$O$_4$ ( Fig.~\ref{fig:7}(e)) exhibit a little wider diffusion path compared to that in Li$_5$FeO$_4$  and Li$_{5}$Fe$_{0.25}$Co$_{0.75}$O$_4$. This is expected to be coming from the higher Li vacancy concentration in the case of  Li$_{4.5}$Fe$_{0.5}$Co$_{0.5}$O$_4$ and lower electrostatic repulsion from the Co atom due to its lower oxidation state in  Li$_{5.5}$Fe$_{0.5}$Co$_{0.5}$O$_4$. However,  in Li$_5$FeO$_4$  and Li$_{5}$Fe$_{0.25}$Co$_{0.75}$O$_4$ the oxidation state of Fe/Co, Li vacancy concentration, and the equilibrium volume of the unit cells are almost the same and hence they exhibit similar Li-ion conduction pathways.

\begin{figure*}[h]
\centering
\includegraphics[scale=1]{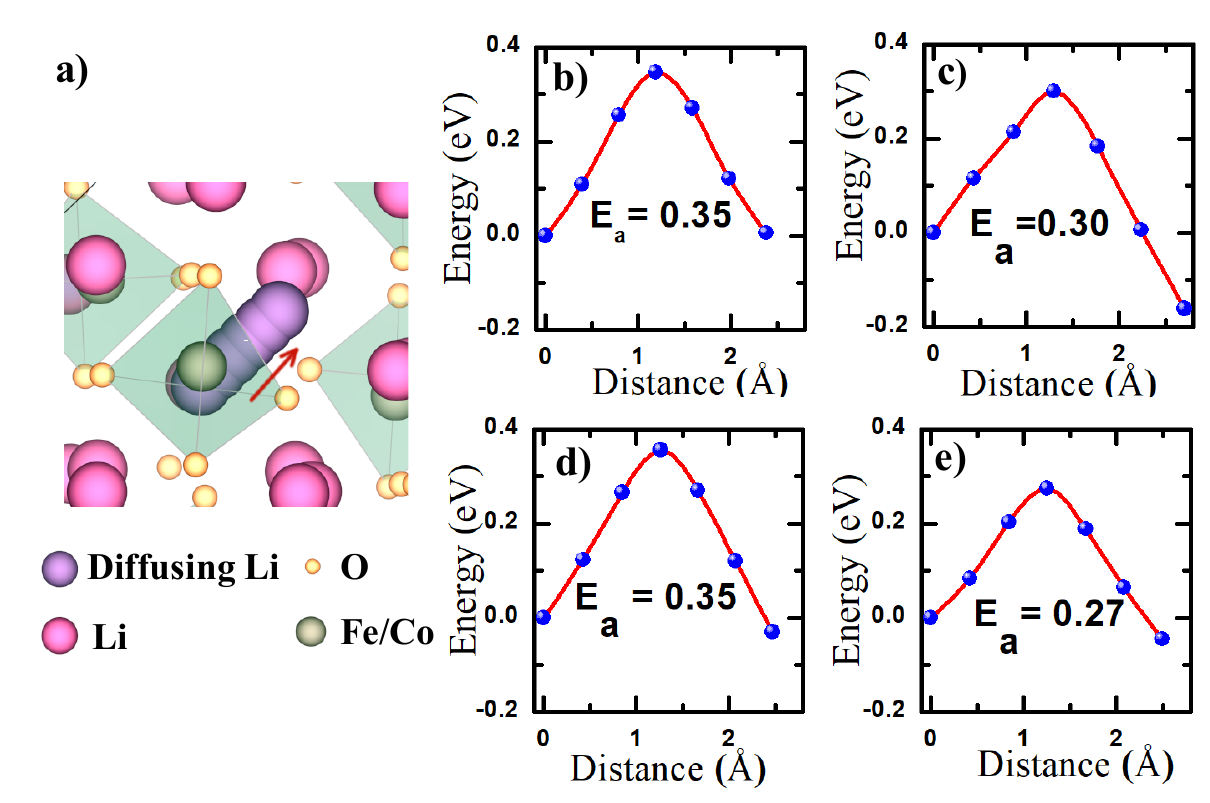}
\caption[c]{ (a) The Li-ion diffusion path in Li$_{y}$Fe$_{(1-x)}$Co$_{x}$O$_4$, (b) diffusion barrier near the Fe site  in  Li$_5$FeO$_4$,  (c) that near  the Co site in  Li$_{5.5}$Fe$_{0.5}$Co$_{0.5}$O$_4$, (d) Li$_{5}$Fe$_{0.25}$Co$_{0.75}$O$_4$, and (e) Li$_{4.5}$Fe$_{0.5}$Co$_{0.5}$O$_4$ obtained from nudged elastic band method}.
\label{fig:8}
\end{figure*}

\par
The BVS map helps to give a qualitative picture of Li-ion diffusion, whereas the nudged elastic band method calculations using VTST scripts~\cite{sheppard2012generalized} and VASP code give a quantitative picture of Li-ion diffusion barrier height. The lower barrier height eases the Li-hop and improves the rate-capacity of Li-ion battery cathodes. In the present study, we have calculated and plotted the diffusion energy for Li-ion diffusion along the migration path in the vicinity of Fe/Co and given in Fig.~\ref{fig:8}. The selected Li-hops used for the CI-NEB calculation are shown in Fig.~\ref{fig:8}(a), where the direction of Li diffusion is indicated by an arrow. The Li ion diffusion barrier near the Fe atom in Li$_5$FeO$_4$  is given in  Fig.~\ref{fig:8}(b). In Fig.~\ref{fig:8}(c)-(d), the Li diffusion barrier near Co atom in Li$_{5.5}$Fe$_{0.5}$Co$_{0.5}$O$_4$,  Li$_{5}$Fe$_{0.25}$Co$_{0.75}$O$_4$, and Li$_{4.5}$Fe$_{0.5}$Co$_{0.5}$O$_4$ are illustrated, respectively. From our calculations, it can be noticed that the barrier height is lower in the case of  Li$_{5.5}$Fe$_{0.5}$Co$_{0.5}$O$_4$ and  Li$_{4.5}$Fe$_{0.5}$Co$_{0.5}$O$_4$  compared to that in Li$_5$FeO$_4$ and  Li$_{5}$Fe$_{0.25}$Co$_{0.75}$O$_4$. This is consistent with the conclusion arrived from our BVS calculations and this lower barrier height of Li diffusion originates from the low valency of Co atom in Li$_{5.5}$Fe$_{0.5}$Co$_{0.5}$O$_4$ and relatively higher Li vacancy concentrations in Li$_{4.5}$Fe$_{0.5}$Co$_{0.5}$O$_4$. It is already  reported~\cite{kang2006electrodes} that the low valency of transition metal cation is beneficial for Li-ion diffusion. It may be noted that the variation in the Li diffusion barrier height with the reaction coordinate along both the directions (forward and backward) in all the considered systems is almost the same except in Li$_{5.5}$Fe$_{0.5}$Co$_{0.5}$O$_4$ as shown in Fig.~\ref{fig:8}. Such change, as shown in  Fig.\ref{fig:8}(c), is expected to come from the different atomic environments of initial and final Li sites, i.e., the initial state has higher energy since the initial Li site is in the vicinity of more Li-ions and experiences more electrostatic force compared to the final Li site in Li$_{5.5}$Fe$_{0.5}$Co$_{0.5}$O$_4$ for the selected hop. In the case of Li$_5$FeO$_4$  and Li$_{5}$Fe$_{0.25}$Co$_{0.75}$O$_4$, the barrier height of the Li diffusion is almost the same due to the same valency of transition metal ions and Li vacancy concentrations. Even though there are some differences in the barrier height values, all these selected systems satisfy the requirement for a commercial cathode material. Also, it is interesting to note that the valence state of  Fe/Co ions and the Li vacancy concentration have a significant effect on the diffusion barrier in the case of Li$_{y}$Fe$_{(1-x)}$Co$_{x}$O$_4$. From this study, we conclude that the substitution of the Co atom at the Fe site in Li$_5$FeO$_4$  does not make a noticeable change in the barrier height of Li diffusion when the Co possesses the same valency as well as the chemical environment as Fe.

\subsubsection{The calculation of diffusivity of Li in Li$_{y}$Fe$_{(1-x)}$Co$_{x}$O$_4$}
\label{The calculation of diffusivity of Li in Li$_{y}$Fe$_{(1-x)}$Co$_{x}$O$_4$}

With the assumption that the diffusion mechanism is independent of temperature, one can use the following Arrhenius law (Equation:\ref{eu_eqn3}) to evaluate the diffusivity at the elevated temperatures.

\begin{equation} \label{eu_eqn3}
 D = d\textsuperscript{2}\nu ~exp\left ( \frac{-E\textsubscript{a}}{k\textsubscript{B}T} \right ),
\end{equation}

where, $D$ is the diffusivity, $d$ is the hopping distance, $\nu$ is the attempt frequency which is assumed~\cite{kang2009first} to be 10$^{13}$, k$_B$ is the Boltzmann constant and T is the temperature. In the present study we have calculated the diffusivity of  Li ion in Li$_{y}$Fe$_{(1-x)}$Co$_{x}$O$_4$ at room temperature (300\,K) and reported it in Table~\ref{tab:8}. The Li-ion diffusivity in all the selected compounds are in the range of electrode materials currently used in commercial applications~\cite{ma2013effect}. Among the selected systems, Li$_{4.5}$Fe$_{0.5}$Co$_{0.5}$O$_4$ exhibits few orders of higher value of Li-ion diffusivity coming from its lower barrier height (0.27 eV). Li$_{5.5}$Fe$_{0.5}$Co$_{0.5}$O$_4$  shows Li-ion diffusivity lower than that of Li$_{4.5}$Fe$_{0.5}$Co$_{0.5}$O$_4$  and higher than that of other systems considered in the present study due to its lower barrier height (0.30 eV) and higher hoping distance (2.70 \AA ). The Li-ion diffusivity of pristine Li$_5$FeO$_4$ is very close to that of Li$_{5}$Fe$_{0.25}$Co$_{0.75}$O$_4$  due to the fact that both materials possess  same barrier height (0.35 eV) and nearly the same hoping distance ( Li$_5$FeO$_4$: 2.38\AA, Li$_{5}$Fe$_{0.25}$Co$_{0.75}$O$_4$ :2.47\AA ). It may be noted from  the Table~\ref{tab:8}  that the barrier height of Li-ion diffusion in   Li$_{y}$Fe$_{(1-x)}$Co$_{x}$O$_4$ governs the Li-ion diffusivity while the hoping distance has a minor impact. 
\begin{table}[]
\centering
\caption{The calculated Li-ion diffusivity in Li$_{y}$Fe$_{(1-x)}$Co$_{x}$O$_4$ at room temperature (300 K) using Arrhenius equation.}
\begin{threeparttable}
\label{tab:8}
\begin{tabularx}{170pt}{ c c }
\hline
Compound &  Diffusivity(cm$^2$ s$^{-1}$)   \\ 
\hline
Li$_5$FeO$_4$  & 7.46 x 10$^{-9}$ \\

Li$_{5.5}$Fe$_{0.5}$Co$_{0.5}$O$_4$  &  6.62 x 10$^{-8}$\\
Li$_{5}$Fe$_{0.25}$Co$_{0.75}$O$_4$  &  7.99 x 10$^{-9}$ \\
Li$_{4.5}$Fe$_{0.5}$Co$_{0.5}$O$_4$  & 1.80 x 10$^{-7}$\\
\hline
\end{tabularx}
\label{tab:8}
\end{threeparttable}
\end{table}

\subsection{The calculation of open circuit voltage in Li$_{y}$Fe$_{(1-x)}$Co$_{x}$O$_4$}
\label{subsection: The calculation and analyses of open circuit voltage in Li$_{y}$Fe$_{(1-x)}$Co$_{x}$O$_4$}

\subsubsection{Calculation of average voltage}
\label{Calculation of average voltage}

\par
The voltage is one of the most critical electrochemical characteristics of cathode materials used in LIBs. The energy density and capacity of cathode materials for batteries are directly related to their voltage. The voltage of a cathode material should be high enough to provide superior energy density. At the same time, it should fall under the stability window of electrolytes used in LIBs~\cite{urban2016computational}. The average intercalation voltage for the Li composition {\it $x_1$} to {\it $x_2$}  can be directly calculated based on the energy difference of initial and final compositions~\cite{ceder1998identification}. If one uses Li$_5$FeO$_4$  as a cathode and Li metal as the anode, the average intercalation voltage is given as:
\begin{equation}\label{eu_eqn4}
   \overline{V}  = -\frac {E^{Li_{x_{1}}FeO_{4}}- E^{Li_{x_{2}}FeO_{4}}- (x_{1} -x_{2}) E^{Li}}{(x_{1} -x_{2})}
\end{equation}
with {\it $x_{1} >  x_{2}$},

where, $\overline{V}$ is the average voltage, $E^{Li_{x_{1}}FeO_{4}}$, $E^{Li_{x_{2}}FeO_{4}}$, and  $E^{Li}$ are the total energies of  $Li_{x_{1}}FeO_{4}$, $Li_{x_{2}}FeO_{4}$, and Li metal, respectively. In the present study we have calculated $\overline{V}$ for the selected systems during delithiation condition using both GGA + {\it U} and SCAN + {\it U} functional. In a previous study  on calculation of $\overline{V}$ for Li$_5$FeO$_4$ with various functionals indicates that the SCAN + {\it U} functional gives the  $\overline{V}$ value in  better agreement with experimental voltage values~\cite{augustine2021ti}. The calculated $\overline{V}$ values of the selected systems in the Li$_{y}$Fe$_{(1-x)}$Co$_{x}$O$_4$ series are given in Table~\ref{tab:9} along with the average voltage value of pristine Li$_5$FeO$_4$ for which experimental voltage measurement is available to compare with.

\begin{table}[]
\centering
\caption{The calculated average voltage (in V) values  for selected  Li$_{y}$Fe$_{(1-x)}$Co$_{x}$O$_4 $ systems using GGA + {\it U} and SCAN + {\it U} functionals. The voltage value obtained from experimental measurement is given in parenthesis}
\begin{threeparttable}
\label{tab:9}
\begin{tabularx}{190pt}{ c c c }
\hline
\multirow{2}{*}{Compound} &  \multicolumn{2}{c}{Average voltage ($\overline{V}$)}  \\\cline{2-3}
  & GGA + {\it U} & SCAN + {\it U}   \\ 
\hline
Li$_5$FeO$_4$  & 3.77 & 4.38   \\
 & (4.7\tnote{{\it a}}) &\\
Li$_{5.5}$Fe$_{0.5}$Co$_{0.5}$O$_4$  & 3.57 & 4.09 \\
Li$_{5}$Fe$_{0.25}$Co$_{0.75}$O$_4$  & 3.63 & 4.20 \\
Li$_{4.5}$Fe$_{0.5}$Co$_{0.5}$O$_4$  & 3.87 & 4.40 \\
\hline
\end{tabularx}
\begin{tablenotes}
\item[{\it a}]{Ref.~\cite{zhan2017enabling}- experimentaly obtained by removing more than 4 Li from formula unit} 
\end{tablenotes}
\label{tab:9}
\end{threeparttable}
\end{table}

From Table~\ref{tab:9}, it is clear that the calculated $\overline{V}$ obtained from SCAN + {\it U} calculation is found to be in good agreement with the experimental value. So, we have used  $\overline{V}$ obtained from SCAN + {\it U} for our analysis. The calculated values of  $\overline{V}$ indicate that the Co substitution at the Fe site in Li$_5$FeO$_4$ does not make much changes in its average voltage. Our calculated   $\overline{V}$  for Li$_{5.5}$Fe$_{0.5}$Co$_{0.5}$O$_4$  obtained from SCAN + {\it U} calculation  exhibits the voltage 0.29\,V less than that of the parent system and on the other hand  that for the  Li$_{4.5}$Fe$_{0.5}$Co$_{0.5}$O$_4$ shows 0.02\,V higher voltage. Apart from these small variations, all these systems are found to have reasonably high voltage and are in the voltage window of commercially available electrolytes suggesting their practical applications.

\subsubsection{Convex hull and voltage profile calculations for Li$_{y}$Fe$_{(1-x)}$Co$_{x}$O$_4 $ systems at 0\,K}
\label{Convex hull and voltage profile calculations}

\par
\begin{figure*}[h]
\centering
\includegraphics[scale=1]{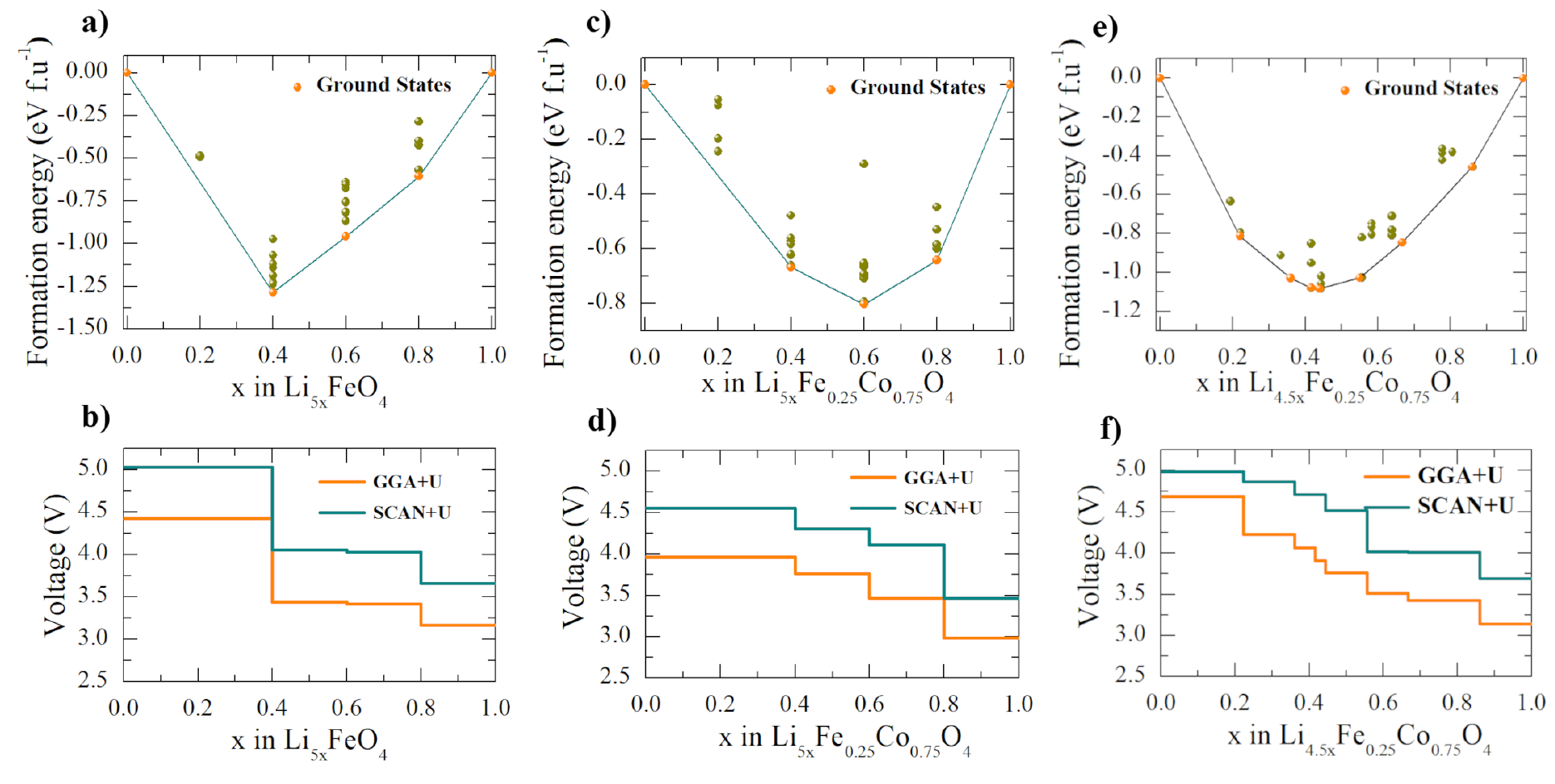}
\caption[c]{The formation energy per formula unit for 32 configurations belonging to six different Li concentrations for (a) Li$_5$FeO$_4$, (c) Li$_{5}$Fe$_{0.25}$Co$_{0.75}$O$_4$  and 15 different Li concentrations for (e) Li$_{4.5}$Fe$_{0.5}$Co$_{0.5}$O$_4$. The olive green and orange colored dots in all the curves correspond to formation energies from DFT calculations and DFT ground states, respectively. The formation energy convex hull connecting the DFT ground states is shown by a blue line. The voltage profile for  (b) Li$_5$FeO$_4$, (d) Li$_{5}$Fe$_{0.25}$Co$_{0.75}$, and (f) Li$_{4.5}$Fe$_{0.5}$Co$_{0.5}$O$_4$) are obtained from GGA + {\it U }and SCAN + {\it U} methods with ground state configurations obtained from the  convex hull.}
\label{fig:9}
\end{figure*}

In order to have  detailed understanding of voltage changes at different stages of delithiation, we have plotted the formation energy convex hull for the complete delithiation process in Li$_5$FeO$_4$, Li$_{5}$Fe$_{0.25}$Co$_{0.75}$O$_4$ and Li$_{4.5}$Fe$_{0.5}$Co$_{0.5}$O$_4$.  It may be noted that the phase stability at different stages of delithiation can be compared using the formation energy with respect to the stable reference phases~\cite{urban2016computational}. We have calculated the formation energy of arbitrary Li concentrations for composition such as Li$_{5x}$FeO$_4$ using the following equation:

\begin{equation}\label{eu_eqn5}
  E_{f}^{Li_{5x}FeO_{4}}=  E_{t}^{Li_{5x}FeO_{4}} - x E_{t}^{Li_{5}FeO_{4}}- (1 -x) E_{t}^{FeO_{4}}, 
\end{equation}

where  $E_{f}^{Li_{5x}FeO_{4}}$ is the formation energy of $Li_{x}FeO_{4}$. $  E_{t}^{Li_{5x}FeO_{4}}$, $E_{t}^{Li_{5}FeO_{4}}$, and $ E_{t}^{FeO_{4}}$ are the total energies of $Li_{5x}FeO_{4}$, $Li_{5}FeO_{4}$, and $ FeO_{4}$, respectively  obtained from {\it ab-initio} calculations. We have taken 32 different configurations for each hull, and the configurations are enumerated using the CASM package while assuming the same occupancy for the same type of Li-atoms (there are five different types of Li atoms in the unit cell of the pristine Li$_5$FeO$_4$). Approaching each Li in  Li$_5$FeO$_4$ as a separate entity will result a vast configurational space since there are 40 Li atoms in the unit cell. This produces 2$^ {40}$ configurations that are in the order of astronomical number (the number of stars in the galaxy)~\cite{turchi2012statics}. Removing symmetrically equivalent structures also leaves out many thousands of configurations, and calculating the total energies of resultant configurations is still computationally expensive. Therefore, in the present study, we have restricted the number of configurations based on atom type and plotted the convex hull as shown in Fig.\ref{fig:9}. The convex hull is formed by connecting the lowest energy configurations (DFT ground states) using a tie line. The slope of the convex hull is directly related to the average voltage~\cite{sudarsanan2021investigation}. In Fig.~\ref{fig:9}(a) and \ref{fig:9}(c),  when the concentration $x = 0.2$, none of the structures falls on the convex hull indicating that none of them are stable at 0\,K. Along with DFT ground states, multiple metastable states fall just above the convex hull in most of the intermediate Li concentrations in all the systems, as shown in \ref{fig:9}. These metastable configurations are important to understanding the voltage profile at elevated temperatures~\cite{chen2018first}.
\par
Armed with the knowledge of intermediate ground state structures from the convex hull, we have calculated the 0\, K voltage profile by calculating the average voltage between adjacent stable phases in the convex hull using equation \ref{eu_eqn4}.  In accordance with the average voltage calculations, we have used both GGA + {\it U} and SCAN + {\it U} functionals to obtain the 0\,K voltage profiles of  Li$_5$FeO$_4$, Li$_{5}$Fe$_{0.25}$Co$_{0.75}$O$_4$, and  Li$_{4.5}$Fe$_{0.5}$Co$_{0.5}$. The calculated voltage profile using GGA + {\it U} and SCAN + {\it U} for Li$_5$FeO$_4$, Li$_{5}$Fe$_{0.25}$Co$_{0.75}$O$_4$, and  Li$_{4.5}$Fe$_{0.5}$Co$_{0.5}$ are given in Fig.~\ref{fig:9} (b), (d), and (f), respectively. From these figures, it is clear that in all these systems, the meta-GGA calculations give a higher voltage value than that from the GGA + {\it U} calculations. Compared with the voltage profile from GGA + {\it U} calculation, that from  SCAN + {\it U} calculations are more reliable, and hence we have used the voltage profile from  SCAN + {\it U} calculations for the following analyses. Depending upon the compositions, the plateaus, steps, and sloping regions in the voltage profiles can vary as found in Fig.~\ref{fig:9} (b), (d), and (f). A plateau usually represents a two-phase region, whereas a sloping region represents a solid solution and a step represents ordered compounds.  In the voltage profile of Li$_5$FeO$_4$, we could see an initial plateau around 3.5 V. This represents the two-phase region (anti-fluorite and distorted rocksalt phase (DRP)) as discussed by Chun Zhan {\it et al.}~\cite{zhan2017enabling} based on experimental studies. In their experimental study, the plateau is found up to the removal of the second Li-atom. However, this is extended up to removing the third Li-ion in our calculation. The extension of the plateau until the removal of the third Li-ion in our calculation in Fig.~\ref{fig:9} (b) may be due to the non-inclusion of DRP structures in the formation energy hull in the present study. Moreover, here the direct comparison with experimental results is not possible since it does not include the oxygen release (0.125 O$_2$ molecules released per electron at 3.5 V as observed experimentally~\cite{zhan2017enabling}), and further, it may be noted that our study is a 0\, K voltage analysis. The 0\, K voltage profile should be considered as an approximation for the finite temperature voltage profile since multiple metastable states in the convex hull will participate in the delithiation process at finite temperatures~\cite{chan2012first}.\\
\par
In Fig.~ \ref{fig:9}(b), there is a significant step at $x = 0.4$ during delithiation. The experimental investigation also indicates the presence of such a step in the voltage profile~\cite{zhan2017enabling}, but in the present study, it is relatively larger with a value of around 1\, V. It may be noted that our voltage profile curves are strictly applicable to 0\, K and also we have considered a limited number of configurations for the generation of the convex hull. Further in the experimental conditions, a small amount of oxygen is released during the delithiation process, which could also influence the step height. Moreover, this difference can also be explained on the basis of crystal field splitting. For the experimental study, Fe ions in the material will be in an octahedral coordination (DRP), while in the present study, we have considered tetrahedral coordination (anti-fluorite). When Fe changes its oxidation states from +3 to +4 during the delithiation process, the electronic configuration of Fe will change from a more stable 3{\it d}$^5$ configuration to a less stable 3{\it d}$^4$ configuration and therefore, increase the Li chemical potential of the cathode. It may be recollected that the equilibrium voltage directly depends on the difference of Li chemical potential between the anode and cathode~\cite{van2020rechargeable} and create the step in the voltage profile as shown in Fig.~\ref{fig:9} (b). During Fe$^{3+}$ to Fe$^{4+}$ oxidation,  Fe loses electron from {\it e$_g$} orbital in octahedral coordination while it loses electron from {\it t$_{2g}$} orbital in a tetrahedrally coordinated Fe. The {\it e$_g$} orbitals in an octahedral configuration have higher energy than {\it t$_{2g}$} orbitals in the tetrahedral coordination. This is expected to be the reason behind the large voltage step in our calculation compared with that present in the experimentally measured voltage profile~\cite{zhan2017enabling}. 
\par
Compared to  Li$_5$FeO$_4$, the Li$_{5}$Fe$_{0.25}$Co$_{0.75}$O$_4$  does not show any plateau in the voltage profile until the removal of the third Li-ion. In Fig.~\ref{fig:9}(d), one can notice a slope like region above $x = 0.4$. It may be noted that the slope region in the experimental voltage profile can be idenftied as small steps in the calculated voltage profile curve due to the discrete number of configurations considered. Moreover, at the finite temperatures,  the steps in the 0\, K voltage profile will become more rounded~\cite{wolverton1998first,yao2018interplay}) and hence the steps will be smeared out as slope. Nevertheless, the experimentally observed slope in the voltage profile is interpreted as originating from a solid solution. Most importantly, one could identify that plateaus and long steps are absent until the removal of the third Li. This reveals that there is no formation of multiple phases until the removal of the third Li.  The bonding analyses in section \ref{Analysis of chemical bonding} and  \ref{Bond strength analysis} clearly indicates that, when Co is in +3 oxidation state the Co$-$O bond is stronger than the Fe$-$O bond. It has also to be noted that the TM$-$O bond is found to be the backbone of structural stability in Li$_{y}$Fe$_{(1-x)}$Co$_{x}$O$_4 $ systems. The voltage decay in  Li$_5$FeO$_4$ was ascribed to the structural phase transition from the defect anti-fluorite phase to the distorted rocksalt phase. Zhan et. al, have revealed that such structural phase transition comes from the migration of transition metal and Li atoms from the tetrahedral site to octahedral sites ~\cite{zhan2017enabling}.  So we deduced from bonding and voltage analyses that, in the case of  Li$_{5}$Fe$_{0.25}$Co$_{0.75}$ the stronger bonding between Co$-$O will prevent the TM migrations from tetrahedralral sites to octahedralral sites and thereby improve structural stability during delithiation and suppresses voltage fade. Along with that, the changes in the electronic configuration also contribute in such a way that the Co atoms in Li$_{5}$Fe$_{0.25}$Co$_{0.75}$O$_4$ can easily oxidize from +3 (3{\it d}$^6$- less stable electronic configuration) to +4 ( 3{\it d}$^5$- more stable electronic configuration) compared to Fe.  These could explain the extra reversible capacity of Co substituted Li$_5$FeO$_4$  mentioned by Imanishi et {\it et al.}~\cite{imanishi2005antifluorite} and show the directions for suppressing the voltage fade present in most of the Li-rich cathodes due to TM migrations. As the TM$-$O bond strengh is found to play a vital role on capacity fade, the Li concentration in Li-rich materials should be properly tuned to get a stronger TM$-$O bond for optimal cycling performance. Compared to the other two systems considered for our voltage profile analysis, Li$_{4.5}$Fe$_{0.5}$Co$_{0.5}$ exhibits a different behavior during delithiation since further oxidation of Co $^{+4}$ is highly unfavorable. From Fig.~\ref{fig:9}(f), one can observe that the delithiation after the first Li-ion removal results in a plateau-like curve corresponding to a two-phase region. The following delithiation exhibits a steep voltage curve denoting a solid solution. On further removal of the Li ions, again the system shows a first-order phase transition as in the case of Li$_5$FeO$_4$ and Li$_{5}$Fe$_{0.25}$Co$_{0.75}$O$_4$. \\ 

\par
From the voltage calculations, it can be deduced that even though there are slight variations in voltage by changing Li and Co concentration, all the considered systems have voltage in the range of promising cathode materials. Even when the average voltages of all these systems are almost the same, the voltage profiles show a different behavior, as explained above. This underlines the dependency of voltage on transition metals, their oxidation states, and their concentration. Among all the considered systems for voltage calculations, Li$_{5}$Fe$_{0.25}$Co$_{0.75}$O$_4$ is found to be the most promising candidate due to the absence of first-order phase transition until the removal of the third Li-ion and expected to provide a reversible capacity of around 513 mAhg$^{-1}$. The delithiation of multiple Li-ions cannot be solely explained on the basis of cation redox with its standard oxidation states [Fe(+2, +3, +4), Co(+2, +3, +4)]. This point towards the presence of anionic redox during higher concentration of Li exchange in Co substituted systems similar to that in pristine Li$_5$FeO$_4$.\\

\subsection{Cationic and anionic oxidation  in Li$_{y}$Fe$_{(1-x)}$Co$_{x}$O$_4 $ during delithiation }
\label{Cationic and anionic oxidation}

\par
\begin{figure}[h]
\centering 
\includegraphics[scale=0.9]{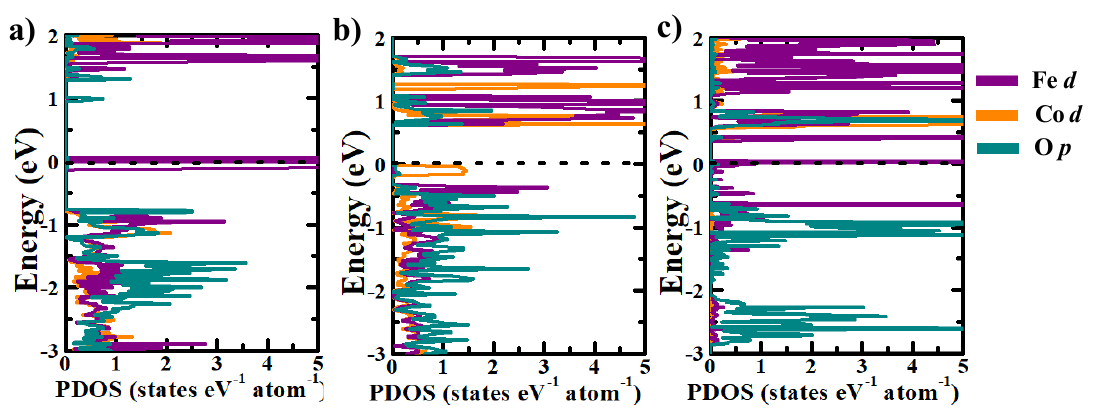}
\caption[c]{The orbital projected partial density of states of Fe-3{\it d}, Co-3{\it d}, and  O-2{\it p} electrons at various stages of delithiation in  Li$_{5x}$Fe$_{0.25}$Co$_{0.75}$O$_4$ (with x = (a) 1.00 , (b) 0.60 ,   (c) 0.00) }
\label{fig:10}
\end{figure}

 
In order to analyze the role of Fe/Co and O atoms in the delithiation process, we have calculated the PDOS of Fe, Co, and O in Li$_{5}$Fe$_{0.25}$Co$_{0.75}$O$_4$ ( Fig.~\ref{fig:10}) and the corresponding Bader charges in Li$_5$FeO$_4$, Li$_{5}$Fe$_{0.25}$Co$_{0.75}$O$_4$ and Li$_{4.5}$Fe$_{0.5}$Co$_{0.5}$O$_4$ (Table~\ref{tab:10}) at different stages of the delithiation process. The PDOS will act as a qualitative measure of the participation of Fe/Co and O atoms in the charge compensation process during delithiation. In the initial stages of delithiation, the presence of a finite amount of DOS of a particular atomic species in the vicinity of the VBM indicates its possible participation in the oxidation process ~\cite{johannes2016oxygen}. Moreover, in the following stages of the delithiation process, the PDOS of a particular atomic species shift from VBM to CBM will confirm its participation in the oxidation process. In the Fig.~\ref{fig:10}, we have shown the orbital projected PDOS of Fe-{\it d}, Co-{\it d}, and O-{\it p} states  in Li$_{5x}$Fe$_{0.25}$Co$_{0.75}$O$_4$ at different stages of delithiation conditions such as $x = (a)1.00, (b) 0.60, (c) 0.00$. From a comparative analysis of  Fig.~\ref{fig:10} (a), (b), and (c), one can deduce that some of the occupied states from  Fe, Co, and O get gradually shifted to the conduction band during delithiation. It indicates that in Li$_{5}$Fe$_{0.25}$Co$_{0.75}$O$_4$ system, the Fe, Co, and O ions undergo oxidation during Li removal. To get a clear picture of the oxidation process during delithiation, we have calculated the Bader charge at Fe, Co, and O sites at various stages of delithiation in Li$_{y}$Fe$_{(1-x)}$Co$_{x}$O$_4 $ and listed them in Table~\ref{tab:10}. From the calculated values of the Bader charges in this table, one can observe that the charge at the Fe site is increasing from 1.50  to 1.61\,e for Li$_{5x}$FeO$_4$ when the $x$ change from 1 to 0.4 and the change in the Bader charge is maximum in the initial stage ( nearly 0.08\,e for $x =1$ to $x = 0.8$ ) of delithiation.  This indicates that the Fe atom gets oxidized from +3 to +4 in these systems during delithiation. However, the Co ions in Li$_{5x}$Fe$_{0.25}$Co$_{0.75}$O$_{4}$ and Li$_{4.5x}$Fe$_{0.5}$Co$_{0.5}$O$_{4}$ behaves differently during delithiation due to different atomic environments of Co in these two systems and also due to their different initial oxidation state. In  Li$_{5x}$Fe$_{0.25}$Co$_{0.75}$O$_{4}$, initially, the Co atom possesses a +3 oxidation state with a Bader charge of 1.32\,e. During the delithiation process, the charge at these Co sites increases and becomes 1.50\,e and then reduced to 1.41\,e when we remove more Li ions,  implying that Co will be oxidized to a +4 oxidation state and slightly reduce its oxidation state again during further delithiation. Whereas, in Li$_{4.5x}$Fe$_{0.5}$Co$_{0.5}$O$_{4}$,  the Co ions will be in a +4 oxidation state initially get slightly decreased during the initial delithiation process. However, the oxidation state increased further during delithiation. It should be noted that the increase in the Bader charge of the Co atom from the initial state to its maximum value is not prominent since the Co atom with a +5 oxidation state is uncommon and highly unstable. Moreover, the PDOS of the initial state, as shown in Fig.~\ref{fig:4}(d),  indicates that the Co-3{\it d} states are 0.5\,eV below the VBM. During delithiation, these Co-3{\it d} states shift close to the VBM and participate in the oxidation process via a small amount of charge transfer.  It can be noted from Table~\ref{tab:10} that the charge at the O sites gets less negative in each delithiation step, and the degree of the change increases with an increase in delithiation, indicating the involvement of O in the oxidation reaction during delithiation. The Bader charge at the O sites in all the systems exhibits a similar trend, confirming the participation of oxygen in the oxidation process during delithiation.

\par
From the analysis of PDOS and Bader charge of  Fe, Co, and O atoms in Li$_{y}$Fe$_{(1-x)}$Co$_{x}$O$_4 $, it can be deduced that both transition metals and O atoms participate in the redox process during Li exchange, i.e., all the selected systems exhibit cationic and anionic oxidation simultaneously during delithiation. While the oxidation of Fe/Co is limited due to their limited preferable oxidation states, the oxidation of O promotes further removal of Li-ions. Nevertheless, careful analysis of the oxygen oxidation reaction is required to confirm whether the O atom undergoes a redox reaction during the lithiation/delithiation process or ends up with oxygen release. So, in the following section, we check the possibility of oxygen release during delithiation in selected systems.\\

\begin{table}[]
\centering
\caption{The calculated Bader charge (in $e$) of transition metals and O atoms during delithiation in Li$_5$FeO$_4$, Li$_{5}$Fe$_{0.25}$Co$_{0.75}$O$_4$, and Li$_{4.5}$Fe$_{0.5}$Co$ _{0.5}$Oi$_{4}$}
\begin{threeparttable}
\label{tab:10}
\begin{tabularx}{205pt}{ c c c c c }
\hline
\multirow{2}{*}{Compound} & \multirow{2}{*}{x} &  \multicolumn{3}{c}{Bader charges (e)}  \\\cline{3-5}
  &  & Fe & O & Co   \\ 
\hline
Li$_{5x}$FeO$_4$ & 1 & 1.50 &  -1.46 & \\
& 0.80 & 1.58 & -1.28 & \\
& 0.60 & 1.60 & -1.07& \\
& 0.40 & 1.61 & -0.85 & \\
& 0.00 & 1.59 & -0.40 & \\
Li$_{5x}$Fe$_{0.25}$Co$_{0.75}$O$_{4}$ & 1 & 1.50 & -1.43 & 1.32  \\
& 0.80 & 1.57  & -1.22 & 1.40 \\
& 0.60 & 1.59 & -1.05 & 1.48  \\
& 0.40 & 1.60 & -0.83 & 1.50 \\
& 0.00 & 1.59 & -0.36 & 1.41\\
Li$_{4.5x}$Fe$_{0.5}$Co$_{0.5}$O$_{4}$ & 1 & 1.50 & -1.36 & 1.46  \\
& 0.86 & 1.57 &  -1.22  & 1.34 \\
& 0.66 & 1.60 & -1.06 & 1.51 \\
& 0.42 & 1.61 & -0.81 & 1.49  \\
& 0.00 & 1.58 & -0.38 & 1.50  \\
\hline
\end{tabularx}
\end{threeparttable}
\end{table}

\subsection{Analysis of oxygen release during delithiation in Li$_{y}$Fe$_{(1-x)}$Co$_{x}$O$_4 $ }
\label{Analysis of oxygen release}

\par
The oxygen released in the cathode material during the delithiation process is detrimental to its structural and electrochemical properties. Also, the oxygen release sometimes even causes thermal runaway of the batteries. Whereas controlling the oxygen oxidation in a  reversible limit provides extra capacities to the cathode materials. As mentioned above, since the Li$_{5x}$FeO$_4$ and its Co substituted derivatives exhibit anionic redox during Li exchange, we have calculated the oxygen release energy at various stages of delithiation condition using equation \ref{eqn:6}. 

\begin{equation}\label{eqn:6}
   \Delta H_{ O\textsubscript{2  }  release} = \frac{E^{Li_{5}FeO_{4-z}}_t+(z/2)E^{O_{2}}_t-E^{Li_{5}FeO_{4}}_t}{z/2}.
\end{equation}

\begin{figure}[h]
\centering
\includegraphics[scale=1]{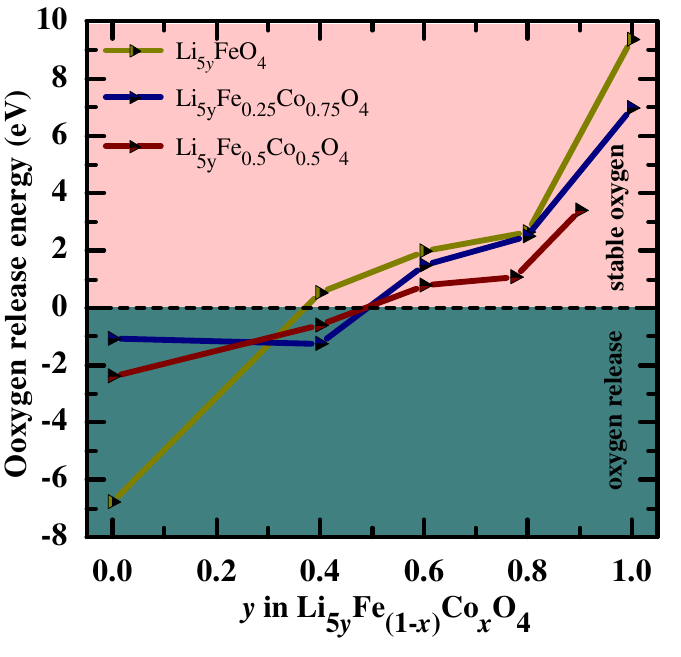}
\caption[c]{The oxygen release energy (eV/O$_2$molecule) of  Li$_5$FeO$_4$, Li$_{5}$Fe$_{0.25}$Co$_{0.75}$O$_4$, and Li$_{4.5}$Fe$_{0.5}$Co$_{0.5}$O$_4$ as a function of Li concentration.}
\label{fig:11}
\end{figure}

Where $ \Delta H_{ O\textsubscript{2  }  release}$ is the oxygen release energy and $E^{Li_{5}FeO_{4-z}}_t$ , $E^{O_{2}}_t$, and $E^{Li_{5}FeO_{4}}_t$ are the total energies of Li\textsubscript{5}FeO\textsubscript{4-z}, O\textsubscript{2} molecule, and Li\textsubscript{5}FeO\textsubscript{4},  respectively. The total energy of the O$_2$ molecule is calculated from the more accurate total energies of H$_2$O and H$_2$ from calculations using the supercell approach, as mentioned elsewhere~\cite{sudarsanan2021investigation}. The calculated oxygen release energy as a function of delithiation in Li$_{y}$Fe$_{(1-x)}$Co$_{x}$O$_4 $  is plotted in Fig.~\ref{fig:11}. A negative value of oxygen release energy indicates the oxygen release during delithiation, while a positive value denotes stable oxygen configuration. From Fig.~\ref{fig:11}, it can be seen that all the selected systems exhibit stable oxygen configurations until the removal of the third Li.  In the parent system, i.e., Li$_5$FeO$_4$, removing the fourth Li-ion leads to a stable oxygen configuration since the oxygen release energy is still positive but close to zero. The oxygen release in the cathode materials is usually initiated by oxygen dimerization (2O$^{2-}$  $\rightarrow$  O$_{2}^{n-}$). When the Li ions are removed from these compounds, the Fe/Co$-$O and O$-$O bond length decreases and gradually leads to oxygen dimerization and oxygen release. According to the Bader charge analysis in the previous section, O atoms are gradually getting oxidized from the beginning of the delithiation process itself.  Moreover, from Bader charge and oxygen release calculations, one can deduce that in all the three selected systems, three Li ions per formula unit can be taken out safely by the cumulative redox of transition metals without any oxygen release, and beyond that, one could expect oxygen release from the cathode that will destabilize the battery.

\section{Conclusion}
\label{Conclusion}

\par

In the present study, we have investigated capacity fade suppression by Co substitution in Li$_5$FeO$_4$  through a systematic analysis of structural, magnetic, thermodynamic, electronic, and electrochemical properties of Co substituted  Li$_5$FeO$_4$ with varying  Li, Fe, and Co concentrations and charge on Co using first-principles calculations with GGA+$U$ and SCAN+ {\it U } functionals. The calculated bandgap values indicate that the partially Co substituted systems exhibit better electronic conductivity than the parent, and fully Co substituted systems. From the detailed analysis of bandgap values evaluated from the density of states, we found that Li$_{5.5}$Fe$_{0.5}$Co$_{0.5}$O$_4$, Li$_{5}$Fe$_{0.25}$Co$_{0.75}$O$_4$, and Li$_{4.5}$Fe$_{0.5}$Co$_{0.5}$O$_4$ are the most promising candidates among all the considered systems, and hence they were selected for further investigations along with the parent material. The oxidation state of Fe and Co ions have been verified using various charge partitioning schemes such as Born effective charge, Bader effective charge, Mulliken charge, and L{\"o}wdin charge population analyses and found that Fe is in +3 oxidation state in all the considered systems, whereas, Co is in +2, +3, and +4 oxidation states in Li$_{5.5}$Fe$_{0.5}$Co$_{0.5}$O$_4$, Li$_{5}$Fe$_{0.25}$Co$_{0.75}$O$_4$ and Li$_{4.5}$Fe$_{0.5}$Co$_{0.5}$O$_4$, respectively. The chemical bonding analyses based on charge density, ELF, and ICOBI indicate that the Li$-$O bond is mainly ionic, whereas the Fe/Co$-$O bonds have an iono-covalent character. From the bond strength calculations, the Fe/Co$-$O bonds are identified as the backbone of structural stability since the Li$-$O interactions are much weaker than the Fe/Co$-$O bonding interactions. This will prevent structural collapse during delithiation. Moreover, the Co$-$O bond strength is found to increase while the charge on the Co atom increases by removing Li-ions adjacent to it. The quantitative and qualitative studies on Li diffusion properties in  Li$_{y}$Fe$_{(1-x)}$Co$_{x}$O$_4 $ have been carried out using the BVS and CI-NEB methods as a function of Li concentration and Co substitution. The BVS method suggests a three-dimensional pathway for long-range Li-ion diffusion in these systems. The CI-NEB calculations and the diffusivity analysis indicate that, both Li$_{4.5}$Fe$_{0.5}$Co$_{0.5}$O$_4$ and  Li$_{5.5}$Fe$_{0.5}$Co$_{0.5}$O$_4$  exhibit prominent Li-ion diffusion characteristics in comparison with Li$_{5}$Fe$_{0.25}$Co$_{0.75}$O$_4$ and  Li$_5$FeO$_4$. From these analyses, we found that the oxidation state of transition metals and Li vacancy concentration were found to play a vital role in Li diffusion kinetics in these materials. All the selected materials are found to have average voltage within the range of commercial cathodes used in LIBs. The calculated 0\, K voltage profiles signify the role of transition metal and its oxidation state on the delithiation mechanism.  Our calculations show that both Li$_5$FeO$_4$ and Li$_{4.5}$Fe$_{0.5}$Co$_{0.5}$O$_4$ exhibit a first-order phase transition in the early stages of the delithiation process itself. But Li$_5$FeO$_4$  substituted with Co in +3 oxidation state suppresses such first-order phase transition. This is attributed to the strong covalent interaction between Co and O in the system and easy oxidation of Co from +3 to +4 state compared to Fe during delithiation. A strong Co$-$O bond suppresses the Co migration from tetrahedral site to octahedral site in Li$_{5}$Fe$_{0.25}$Co$_{0.75}$O$_4$ and thereby the phase transition and capacity fade.  So, one can take out three Li-ions from Li$_{5}$Fe$_{0.25}$Co$_{0.75}$O$_4$  without phase transition and oxygen release and provide a reversible capacity of around 513 mAh$^{-1}$.  The present study shows that the Fe/Co$-$O bonds and the oxidation state of  Co play a vital role in the performance of Li-rich Li$_{y}$Fe$_{(1-x)}$Co$_{x}$O$_4 $ cathodes.  Also, our bond strength and voltage profile analyses indicate that it is important to carefully adjust the Li concentration in Li-rich cathodes as it directly influences the TM$-$O covalency and thereby the cycling performance of the materials.  The simultaneous cationic and anionic redox reaction is identified as the electrochemical reaction behind the delithiation process in Co substituted system as in the pristine Li$_5$FeO$_4$. Moreover, there will not be any oxygen release in all the selected systems until removing the third Li-ion per formula unit. Our investigations prove that a tuned Co substitution will enhance the structural, electronic, and electrochemical properties and suppress the initial capacity fade of the pristine Li$_5$FeO$_4$ and make the material more interesting for practical applications.

\section*{Conflicts of interest}
There are no conflicts to declare.

\section*{Acknowledgements}
The authors are thankful for the fruitful discussion and helpful comments from Helmer Fjellvåg and Alok Mani Tripathi from the University of Oslo, Norway. This work is benefited from the student exchange Indo Norwegian Collaborative Program, "Theoretical and Experimental Studies on Functional Materials (TESFun)," funded by the UTFORSK program of the Centre for Internationalizing Education (SIU) Norway. Computer resources are provided by the SCANMAT center, Central University of Tamil Nadu, India.



\balance


\bibliography{manuscript} 
\bibliographystyle{manuscript} 

\end{document}